# Taking a Closer Look at Warnings Generated by PMD and SonarQube, their Rules and Compliance to Established Coding Standards

Lakmal Deshapriya*, Sherlock A. Licorish, Brendon J. Woodford

*School of Computing*
*University of Otago,*
*Dunedin, New Zealand*
lakmal.deshapriya@postgrad.otago.ac.nz, sherlock.licorish@otago.ac.nz,
brendon.woodford@otago.ac.nz

***Abstract***

***Context:*** *Static code analysis (SCA) tools play a vital role in software development, reducing the cost and time required for code reviews. However, high false-positive and false-negative rates are reported for the best tools in the community. Accordingly, studies often aim to develop datasets for learning SCA warning patterns to reduce false results. These datasets are meant to possess high-quality and high-volume in covering the full range of faults/rules that typically result in false warnings and be compliant with established coding standards. However, existing studies have not utilised such datasets or identified the breadth of rules that are prone to false positives and their compliance to coding standards.* ***Objectives****: We analysed code from Stack Overflow and Apache Tomcat to capture variations in code length and style in detecting false-positive warnings from best-performing tools PMD and SonarQube, addressing this gap.* ***Method****: In deriving false-positive warnings, outcomes from the tools were labelled using established coding standards. Deeper analyses were then conducted to identify the rules that are prone to false-positives, reasons for these, and agreement/gaps between SCA rules and established standards.* ***Results****: Among our main outcomes, we observe that only a few SCA rules generate false positives, ranging from 4.64% to 18.45% across four datasets. Additionally, eliminating rules that contradict established standards significantly reduce the false-positive rate. Additionally, our findings reveal discrepancies between tools and established standards.* ***Conclusion****: Given the evidence established in this study, we recommend further investigations into gaps between tools and established standards, including the use of machine learning approaches to annotate larger datasets.*

## 1 Introduction

Detecting and fixing bugs have become an essential day-to-day task of the software engineering process, and researchers have found that software engineers use more than half of their time on improving software (e.g., fixing bugs and refactoring source code) [1]. Static Code Analysis (SCA) tools were introduced to identify common quality issues in source code [2], in supporting developers' effort aimed at delivering high quality code. However, due to the massive amount of warnings generated by SCA tools, similar to the examples shown in Figure 1, the number of false positive warnings[1] is also considerable, making it difficult to identify actual warnings[2] creating a challenging task [3-5]. The SCA tools aim to identify as many code quality violations as possible, in trying to minimise false negatives [6-8]. Consequently, SCA tools generate many false alarms (false positives), wasting developers' time figuring out the actionable code quality violations among the large volume of warnings generated [3-5]. Therefore, identifying actual code quality violations suggested by SCA tools is essential to the software engineering Practitioner Community.

* Corresponding author

```
public class arrayformat {

   public static void main(String[] s)
   {
         int[] array = {1,2,3,4,5};

         int temp = array[4];

         for(int i=array.length-1;i>0;i--)
         {
            array[i]=array[i-1]; // issue - AvoidArrayLoops
         }

         array[0]= temp;

         for(int i=0;i<array.length;i++)
         {
            System.out.println(array[i]);
         }

   }
}
```

PMD – True positive warning

```
public class C148203{
public interface CompositeIdEntity {

    long getIdA();

    long getIdB();  //issue CommentRequired

 }
```

PMD – False positive warning

```
public class C270994{
public static void main(String[] args) {
   int income = 1000000;  // Sample test amount of income, should
result in 327,683.50 in taxes.
   double tax = 0;
   if (income > 0) {
      tax = income * 0.10;
      tax += (Math.max(0, income-8350)) * 0.05; // issue Magic
numbers should not be used

      tax += (Math.max(0, income-33950)) * 0.10;
      tax += (Math.max(0, income-82250)) * 0.03;
      tax += (Math.max(0, income-171550)) * 0.05;
      tax += (Math.max(0, income-372950)) * 0.02;
   }
   System.out.println("Tax: "  + tax);
}

}
```

SonarQube – True positive warning

```
      int[][] one =
     { { 1, 2, 3 },  // issue Magic numbers should not be used
       { 4, 5, 6 } };
```

SonarQube – False positive warning

Figure 1 True positive and false positive warning examples for PMD and SonarQube

Efforts are being made to identify actual warnings among the warnings generated by SCA tools [9, 10]. However, these studies have been based on unreliable datasets [11]. Automatically annotated datasets, artificially generated datasets, and datasets with limited rules or warning types are regarded as unreliable or incomplete [11]. In this regard, Hegedus and Ferenc [10] annotated a set of warnings generated using SonarQube. They annotated these warnings, analysing “//NOSONAR” comments in the source code, which developers use to disregard a SonarQube warning suggested for the same line. However, this approach can be unreliable, as developers may apply the “//NOSONAR” comments based on personal judgements rather than following any accepted standards or guidelines. Furthermore, the authors have not validated the dataset, and as a result, there can be incorrect labels in the dataset. Therefore, the dataset is not trustworthy for further analysis.

[1] In this study, a warning denotes the recommendation provided by the SCA tool through source code analysis. "Code Quality violation" and "SCA Alarms" are also used interchangeably to refer to the same concept.
[2] In this study, an actual warning denotes a true positive warning generated by SCA tools.

On the other hand, the Juliet vulnerability dataset is an artificially generated dataset [12]. This synthetic dataset was created using predefined patterns, which caused a lack of diversity in the data, even though the dataset size was large [11]. Consequently, models generated using this dataset might produce promising results. However, their generalizability will be low, as there are only a few unique patterns in the dataset [13]. In another work, Li et al. [14] used a Use Before Initialization (UBI) bug dataset to experiment with ChatGPT[3]. The dataset had 20 items, and these authors evaluated whether the bug was real or false positive using ChatGPT. As they used only UBI bugs for their study, the results produced by this study cannot be generalised.

Correspondingly, according to Meldrum et al. [2], SCA rules/warning types should be classified based on a range of source code quality aspects: 1. Reliability and conformance to programming rules, 2. Readability, 3. Performance, and 4. Security. However, existing studies focus only on a few aspects of source code quality to train Machine Learning (ML) models to identify actionable warnings among the large set of warnings typically generated by SCA tools [14-16]. In fact, relying on a few aspects of source code quality does not fully encompass the overall quality of source codes, and the results generated by such studies cannot be generalised to the wide breadth of source code quality indicators [2].

The limitations of automatically generated datasets, synthetic datasets, and datasets with limited rules or warning types undermine the results produced by those studies, making outcomes unreliable. In addition, the poor quality of the existing annotated datasets makes it significantly challenging to build good-performing ML models to identify actionable warnings as models trained with inaccurate datasets tend to generate inadequate outcomes [11-13]. Among the poor quality of existing annotated datasets, incorrect labels, duplicate data points, and a lack of diversity of data points in synthetic datasets are significant [11].

Therefore, to address these limitations, we created four manually annotated sample datasets with true positive and false positive warnings covering a wide range of source code quality indicators that facilitate analyses to understand false positive patterns better. These datasets were used to explore the SCA rules that are prone to false positives, reasons for the reported false positives and agreement/*gaps* between SCA rules and established coding standards. Our study makes the following contributions for academia and the software engineering Practitioner Community.

1. We analysed current software standards and identified their application to rules that generated warnings for PMD and SonarQube. This mapping provides developers clearer insight into the connection between these tools and industry standards, also outlining *gaps* between SCA rules and defined standards.
2. Unlike in existing studies, defined standards were used to annotate warnings. Four datasets were labelled manually, providing a testbed for future studies.

[3] https://chatgpt.com/

3. We identified SCA rules that are prone to false positives and examined the reasons for false positives.

The remainder of this study is organised as follows. Section 2 describes the background and related studies. In Section 3, the methodology is introduced and discussed, and Section 4 provides the results. Section 5 discusses the findings of our work, while threats to the outcomes are evaluated in Section 6. Finally, the conclusion and suggested future works are discussed in Section 7. We provide a replication package, which consists of the datasets generated from this study, justifications for decisions, coding rules and standards, and appendices. This will be useful for those interested in further examining our research methodology and performing replication studies (refer to Section 8) [17].

# 2 Background and Related Studies

In existing studies, several approaches were taken to create datasets, including real-world program vulnerability datasets and artificially synthesised vulnerability datasets [13]. In this section, the approaches to creating real-world program vulnerability datasets are reviewed, setting the tone for a review into the strategies for creating artificially synthesised vulnerability datasets. Then, the datasets used in existing studies for ML model building to identify false positive warnings are reviewed, where gaps are synthesised.

## 2.1 Vulnerability datasets

Real-world source codes from GitHub repositories were used to create warning datasets, which generate large amounts of datasets mostly using automatic labelling methodologies, which are reviewed in this sub-section. Zheng et al. [18] introduced a methodology to label SCA warnings using differential analysis of bug fixes commits. They created a dataset called D2A with more than 1.3 million examples using the Infer SCA tool to analyse source codes from multiple C/C++ open-source projects. To annotate the warnings in the dataset, they analysed the commit histories (both the previous and subsequent versions of bug fixes) to determine whether the reported warning in the previous version had been resolved in the subsequent version. If the warning was resolved, it was assumed to be an actual (true positive) warning, otherwise, it was deemed a false positive warning. However, according to Croft et al. [11], the quality of the D2A dataset is poor due to its high volume of duplicate data points; consequently, models developed with this dataset may exhibit strong validation results, as there is an increased likelihood of overlapping data points in both the training and validation sets [11]. As a result, the model can exhibit signs of overfitting to the training data, leading it to predict identical labels for the validation dataset. This behaviour, where the model essentially remembers the training examples rather than learning generalisable patterns, artificially inflates its apparent accuracy on the validation dataset.

Cabral et al. [19] introduced the RVprio to rank runtime verification violations based on the likelihood of a code violation being a true bug. They claimed that RVprio was the first and most comparable tool for ranking runtime verification violations. However, the study did not label the dataset but instead predicted the likelihood value. In another study, Aladics et al. [20] introduced a method based on the SZZ [21] to create a dataset of vulnerability-introducing commits as oppose to existing datasets focussed on vulnerability-fixing commits. The dataset consisted of the vulnerability introducing commits with the details of the vulnerability as well. This dataset was created using an automated procedure, and only a sample of the created dataset was validated manually. Consequently, the reliability of the dataset is questionable. Pereira et al. [22] created a dataset of vulnerabilities for C/C++ using five open-source projects by analysing the commit history. The dataset contained 5,214 data points. As the study did not specify whether the generated dataset was manually verified, the dataset is deemed to be unreliable. Chen et al. [23] also created a similar dataset of vulnerabilities called DIVERSEVUL, which consisted of 18,945 vulnerable functions. The dataset was used to identify vulnerabilities in software source code employing 11 ML models.

Furthermore, Wang et al. [24] created a similar dataset of vulnerabilities called ReposVul using 1,491 projects. However, these datasets were not labelled in terms of whether the vulnerabilities were true positives or false positives. Instead, these datasets were labelled automatically, primarily by analysing commit histories. According to Croft et al. [11], such datasets are of low quality and contain incorrect label information. Consequently, artificially synthesised datasets were introduced to address this issue. For instance, the SAMATE project has developed the Juliet vulnerability dataset [12]. The Juliet dataset is used in multiple studies to evaluate ML models; however, the advantages of using artificially synthesised vulnerability datasets come with a cost [11]. Despite the large dataset size, the synthetic dataset was created using predefined patterns, resulting in a lack of diversity in the data [11]. Consequently, ML models generated from this dataset may yield promising results. However, their generalisability is likely to be low, as there are only a limited number of unique patterns in the dataset [13].

## 2.2 Datasets used in ML model building

This sub-section discusses warning datasets used in ML model-building tasks from recent studies. Tan and Tian [15] trained various ML models, including Bidirectional Gated Recurrent Unit Neural Network (BGRU), Convolutional Neural Networks (CNN), and Multilayer Perceptron (MLP), to classify bugs using a real bug database called Defects4J [25]. Furthermore, Yang et al. [26] used three datasets: a bug dataset created by Yang et al. [26], Defects4J [25], and a dataset created by Liu et al. [27], which is an annotated warning dataset derived from commit history analysis. Furthermore, Yerramreddy et al. [28] used four datasets (three Java and one C), created from the analysis results of the JBMC and CBMC SCA tools. The four datasets comprised one real-world dataset for Java and three synthetic datasets. The real-world dataset is manually labelled. However, all the datasets utilised by Yerramreddy et al. [28] merely addressed certain aspects of source code quality, such as vulnerabilities and bugs.

Similarly, Li et al. [14] used a UBI bug dataset, which consists of 20 data items, to experiment with ChatGPT in order to assess whether the bugs were real or false positives. Nguyen et al. [16] used the Juliet test suite (for C/C++) to train ML models. The dataset is a collection of test cases organised under 118 different CWEs[4]. Nagaraj et al. [29] created a security bug dataset by analysing 2,740 Java files using the FindSecBugs SCA tool. However, the dataset did not provide details on whether the bug was real or synthesised. Kharkar et al. [30] created a data set of 539 null dereference warnings. The dataset was generated by analysing seven Java repositories using the Infer SCA tool and was labelled by experienced software developers.

Furthermore, Hegedus and Ferenc [10] created a dataset analysing 9,958 different open-source Java projects hosted on GitHub. The dataset consisted of 224,484 sample data points, which include true positive (actionable) and false positive (explicitly discarded) warnings. To compile the dataset, these authors employed SonarQube to analyse the projects and pinpoint the code violations warnings. However, the study depends on "//NOSONAR" comments within the source code, where developers utilise the comment to disregard a sonar warning suggested for the same line. The approach may be unreliable, as it does not consider whether the developer has ignored the warning in accordance with any accepted standard or guideline. Furthermore, the dataset was neither manually verified by the authors nor used in any other existing studies. Consequently, the dataset is deemed untested for further analysis. Vu and Vo [31] used a dataset for their study that was also used by Ngo et al. [32] consisting of 6,620 warnings from 10 popular open-source C/C++ programs. The dataset comprised two categories of vulnerability types: Buffer Overflow and Null Pointer Dereference. Additionally, Tanwar et al. [9] created a dataset of 10 CWE categories and labelled them using commit history analysis. Koc et al. [33] used two datasets to train their models. The first is the OWASP benchmark, and the second is a dataset of warnings generated by the FindSecBugs SCA tool for 14 real-world programs. The first dataset is synthetic, while the second consists of bugs labelled manually. Furthermore, Lee et al. [34] employed 9,871 alarms, which were either resolved or labelled as false positives at Samsung Electronics. The dataset included warnings from in-house static analysis checkers, comprising six types such as Resource Handle Leak, Double Free, and Null Pointer Dereference After Null.

[4] https://samate.nist.gov/SARD/test-suites/112

Evaluating the outcomes above, the quality of the datasets utilised to build ML models for identifying false positive SCA warnings in previous studies was inadequate. Several studies employed synthetic datasets with limited diversity [16, 33], while others utilised automatically generated datasets that lack manual validation [10, 26]. Moreover, some datasets used to train ML models were not diverse; they involved only a limited range of warnings and did not cover a wide range of source code quality dimensions [2]. As a result, the reliability of the findings produced by existing studies remains questionable. Therefore, it is emphasised that constructing ML models using manually annotated datasets that encompass various aspects of source code quality has the potential to yield generalised results, which is the focus of this study.

## 2.3 SCA tools' rules, the applicable standards for these rules and the gaps between SCA tools' rules and established standards

This subsection explored the efforts made to identify the rules of SCA tools, the standards applicable to these rules, and the gaps between the rules of SCA tools and established standards. Fatima et al. [35] conducted a comparative study on a few SCA tools that are used to analyse C/C++ source code. They evaluated those SCA tools based on twenty-eight checks to rank tools and identify the best SCA tools. Although they do not identify all the SCA tool's rules, they have identified that twenty-eight checks are available in the tool as rules. However, they only evaluated some SCA tools for C/C++ source code and did not check the rules against any source code quality standards. Additionally, no false positive checks were carried out in this study. Ramler et al. [36] identified 42 PMD rules which is related to Java Junit test cases. However, they did not identify any possible source code quality standard or false positives generated by PMD rules. Ashfaq et al. [37] conducted a comparative analysis on Java SCA tools, including PMD. They identified supported rule types (e.g., Style, General, Concurrency, etc.) from each SCA tool. However, they did not identify specific rules that SCA tools supported, nor did they establish any potential source code quality standards relevant to the SCA tools' rules.

Novak and Krajnc [38] created a taxonomy of SCA tools using several attributes of SCA tools, including rule types that the SCA tools supported. However, they did not study the individual SCA rule level and did not consider whether these rules are aligned with source code quality standards. Lenarduzzi et al. [39] identified the rule types (e.g. Syntax, Bugs, etc) each SCA tool (including PMD and SonarQube) generated warnings for, the number of rules generated warnings, and the number of warnings generated. They identified the top ten generated warnings from SCA rules.

Charoenwet et al. [40] studied eight CWE[5] vulnerabilities using several SCA tools, including CodeChecker, Codeql, and Flawfinder. However, they evaluated only a few types of vulnerabilities. A similar study was conducted by Ozturk et al. [41] using the top 10 OWASP[6] vulnerabilities. Alqaradaghi and Kozsik [42] evaluated several Java SCA tools against 70 CWEs, validating these CWEs using OWASP. However, they do not conduct a comprehensive analysis of all the SCA rules available in the SCA tools and their applicable standards.

Software quality models, such as ISO/IEC 9126 and ISO/IEC 25010, are defined to evaluate the quality of software [43, 44]. Some efforts were taken to map the quality characteristics mentioned in these quality models with the SCA rules to classify these rules. Forouzani et al. [44] mapped PMD rules with the relevant quality characteristics according to the definition of quality characteristics from ISO/IEC 25010. Some of the quality characteristics mentioned in the software quality models are Performance, Reliability, Security and Maintainability, and these characteristics are further divided into subcategories such as Availability, Fault Tolerance and Compliance [43, 44]. Bánsághi et al. [43] conducted a similar work for PMD and FxCop SCA tools' rules with both ISO/IEC 9126 and ISO/IEC 25010. However, these works did not assess the validity of SCA tools' rules in terms of quality, and they categorised rules based on the quality characteristics defined in the standard.

According to the existing studies discussed in this sub-section, several efforts have been made to identify rules available in SCA tools. However, these efforts were not comprehensive as they did not identify all the rules available in the SCA tools and the established standards applicable to those rules. Although several authors have utilised software quality models to categorise the rules available in SCA tools, none of the studies identify all the rules in these tools, applicable standards for the rules, or gaps between these rules and established standards.

## 2.4 Summary

Existing studies have developed ML models using datasets created through various methodologies. Some datasets were collected and annotated manually [33, 43], while others were gathered and annotated using commit history from GitHub [10, 18, 20]. Certain datasets were also generated using real-world code repositories [10, 18, 20]. However, some datasets were developed using synthetic code databases, such as OWASP benchmarks and Juliet test suites, offering distinct advantages and disadvantages compared to other types of datasets [12]. Furthermore, datasets discussed in this section are summarised with all the available details in Table 1.

---

[5] These CWEs are maintained by Mitra[5], and we selected this as a source code quality standard in the section 3.4.

[6] OWASP was also identified as a source code quality standard in section 3.4.

The main benefit of automated labelling systems is possessing a substantial quantity of data [13]. However, the datasets generated through automatic methods, such as analysing GitHub commit history, experienced quality issues, whereas those created using synthetic code datasets lacked data diversity [11, 13]. Additionally, recent studies have shown that ML models were constructed using only a limited number of code quality violation types; primarily, only warnings related to bugs were utilised in the creation of these models [14, 15]. However, several dimensions of source code quality recommended by previous work [2] were not addressed in the existing studies, leading to a lack of generalisability to the many aspects of code quality, which needs investigation.

Moreover, several studies were conducted to identify rules available in SCA tools [35, 36]. However, these studies did not comprehensively study on all the rules available in SCA tools. Furthermore, only a few studies tried to map the available rules with CWE and OWASP coding standards [40-42]. Several studies mapped quality characteristics of software quality models, such as ISO/IEC 9126 and ISO/IEC 25010, with rules of SCA tools [43, 44]. However, the purpose of these models is to categorise the available rules according to the quality characteristics.

To address the limitations, we manually created four vulnerability datasets, that cover various aspects of source code quality, as explained in Section 3.4. We annotated samples from each vulnerability dataset according to established standards identified in Section 3.4 to determine whether those warnings were true positives or false positives. These four vulnerability datasets were created using two publicly available code datasets and two popular SCA tools, as outlined in the Section 3.6.

Table 1 Summary of the datasets used in existing studies

| Author | Dataset name | Labelling method | Type of the Dataset | Number of examples | Programming Language |
|---|---|---|---|---|---|
| Zheng et al. [18] | D2A | Automatically | Real-world program vulnerability datasets | Over 1.3 million examples | C/C++ |
| Aladics et al. [20] | | Automatically | Real-world program vulnerability datasets | 564 | Java |
| Pereira et al. [22] | | Automatically | Real-world program vulnerability datasets | 5,214 | C/C++ |
| Chen et al. [23] | DIVERSEVUL | Automatically | Real-world program vulnerability datasets | 18,945 | C/C++ |
| Wang et al. [24] | ReposVul | Automatically | Real-world program vulnerability datasets | 6,134 | C/C++, Java, Python |
| Boland et al. [12] | Juliet vulnerability dataset | Manually | Artificially synthesised vulnerability datasets | over 81,000 | C/C++ and Java |
| Just et al. [25] | Defects4J | Automatically | Real-world program vulnerability datasets | 854 | Java |
| Yang et al. [26] | | Manually | Real-world program vulnerability datasets | 74,859 | Java |
| Liu et al. [27], | | Automatically | Real-world program vulnerability datasets | 16,918,530 | Java |
| Kharkar et al. [30] | | Manually | real-world program vulnerability datasets | 539 | Java |
| Li et al. [14] | | Manually | real-world program vulnerability datasets | 20 | |
| Nagaraj et al. [29 | | | | | Java |
| Hegedus and Ferenc [10] | | Manually | real-world program vulnerability datasets | 224,484 | Java |
| owasp.org [7] | OWASP | Manually | artificially synthesised vulnerability datasets | 21,041 in v1.1 (Java)<br>2,740 in v1.2 (Java) | Java and Python |

[7] https://github.com/OWASP-Benchmark/BenchmarkJava

| | | | | | |
|---|---|---|---|---|---|
| | | | | 1,243 (Python) | |
| Lee et al. [34] | | Automatically | Real-world program vulnerability datasets | 9,871 | C/C++ |

# 3 Methodology

This section outlines the research questions of our study, and the methodology employed to create the datasets and analyse the data. In reviewing existing studies in Section 2, it is observed that there were no studies that examined static analysis tools (including PMD and SonarQube) in terms of their rules, the applicable standards for these rules, false positive warnings generated by these tools and the gaps between SCA tools' rules and established standards. Furthermore, existing studies employed datasets of low quality and limited diversity, rendering the outputs generated by these studies unreliable and ungeneralizable. Thus, four research questions were formulated to address these gaps, and four datasets were created to address those research questions, as detailed below.

## 3.1 Research questions

SCA tools check common software source code defects, producing many warnings depending on the source code size. Developers use a large amount of their development effort to address these warnings, and the complexity of that task increases as SCA tools also produce false alarms [1]. SCA tools check source code using defined rules that evolve over time[8,9]. Developers of SCA tools create, modify, or remove existing rules based on current requirements by continuously evaluating changes in the software engineering industry[10]. In this regard, SCA tools often introduce new rules to detect source code defects while retiring some existing rules when they are no longer relevant. However, limited previous studies have investigated the rules *that are prone to false positives* in SCA tools and established standards relevant to them. Consequently, our first Research Question (RQ) addresses this opportunity:

**RQ1: What are the rules and standards of SCA tools that generate warnings for specific source code datasets?**

*RQ1 was created to identify the current rules available that generated warnings for our source code datasets, in particular SCA tools (PMD and SonarQube), as well as the related established standards. In addition, this research question allows us to evaluate the applicability of established standards to SCA rules and the categories of rules which report warnings for our datasets.*

Existing studies have utilised several approaches to develop SCA warning datasets, including automatic annotation methods [10, 18] and the use of synthetic code datasets like OWASP benchmarks and Juliet test suites [12]. However, these methodologies generate low-quality datasets, and it is recommended that researcher create SCA warning datasets by manually annotating them [11]. Therefore, we have created four datasets through manual annotation following established standards. We observed that only a few SCA rules produced false positive

[8] https://docs.sonarsource.com/sonarqube/latest/user-guide/rules/overview/
[9] https://pmd.github.io/pmd/
[10] https://docs.sonarsource.com/sonarqube/latest/user-guide/rules/overview/

warnings, and it is crucial to identify these rules, as none of the existing studies have focused on this gap. Thus, the second research question is dedicated to this opportunity:

**RQ2: Which SCA rules are prone to false positives?**

*RQ2 was created to identify the rules that generate a significant amount of false positive violations. Identifying those rules makes fixing them possible and reduces the overall false positive rate of the SCA tool, thereby reducing practitioners' workload to sift through false positives.*

As with the identification of false positives, recognizing the causes of false positive warnings generated by SCA rules is essential. However, limited current studies investigate the reasons behind false warnings identified by SCA tools. Therefore, the third research question addresses this opportunity:

**RQ3: What are the reasons for reported false positives by SCA tools?**

*RQ3 was developed to identify the reasons for generating false positives for each rule in SCA tools. Understanding the causes of false positive warnings simplifies the process of recognising these warnings in an unlabelled warning dataset. It also provides insights for SCA Tool Developers to enhance the rules that generate a substantial number of false positive warnings, reducing practitioners' overhead.*

Previous studies have not used established standard/s to annotate SCA warnings. Consequently, no potential gaps have been assessed between the rules of SCA tools and established standards. This gap is addressed by our fourth research question:

**RQ4: Are there any gaps in code violation classification between selected SCA tools and established standards?**

*RQ4 was created to understand discrepancies between the description of SCA rules and established standards. There can be mismatches and gaps between SCA rules and established coding standards. The gaps may arise mainly because SCA tools have defined their rules based on the experience of the software engineering community and their preference for coding styles[11]. In contrast, the established coding standards are factually defined, with firm evidence from standard organisations [45, 46]. Additionally, SCA tools may violate their own standards if they wrongly identify code violations – we investigate this issue.*

## 3.2 SCA tool selection

A systematic mapping study incorporating grey literature was conducted using existing research published between 2013 and 2025 on SCA tools, and 52 blog posts that were returned from Google search results. The study aimed to identify popular SCA tools within the software engineering community and academia. Figure 2 is extracted from the systematic mapping study, where it is shown that the most popular SCA tools used to analyse Java source codes are SonarQube and PMD in academia and industry. Accordingly, for our analysis, we used PMD version 6.55.0 and SonarQube version 10.2.1 to analyse Stack Overflow and Apache Tomcat datasets, which are described in Section 3.3, using these SCA tools to generate warning suggestions.

[11] https://community.sonarsource.com/c/clean-code/43

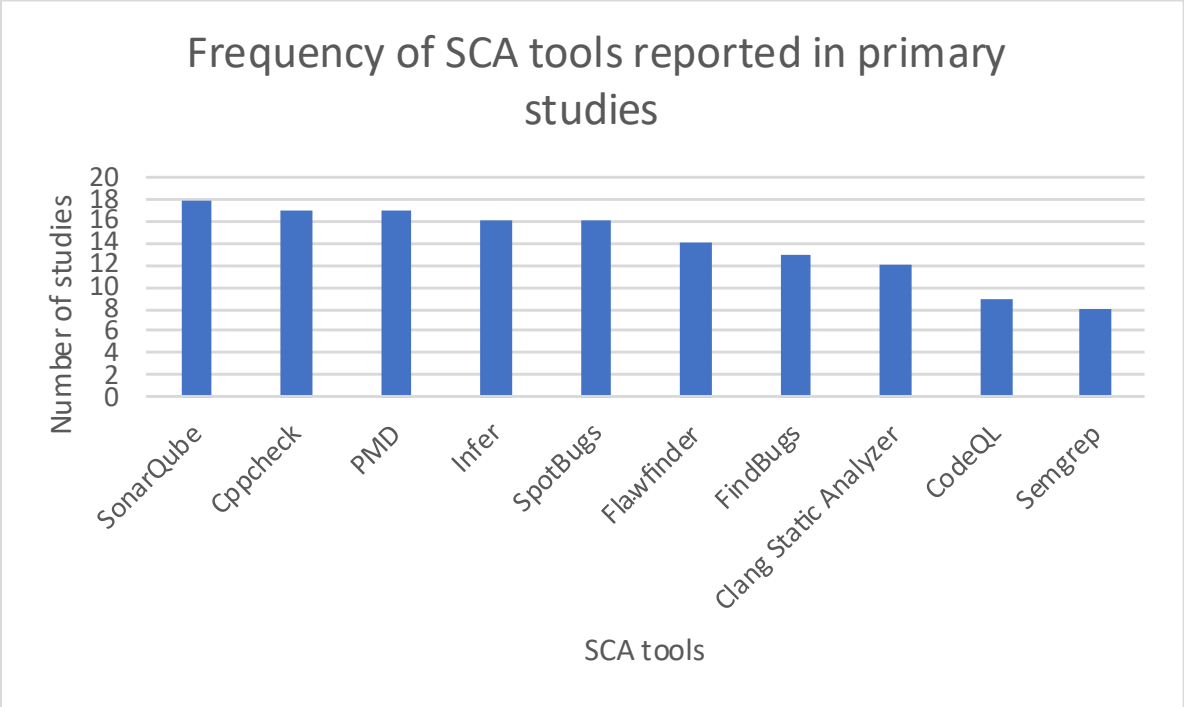

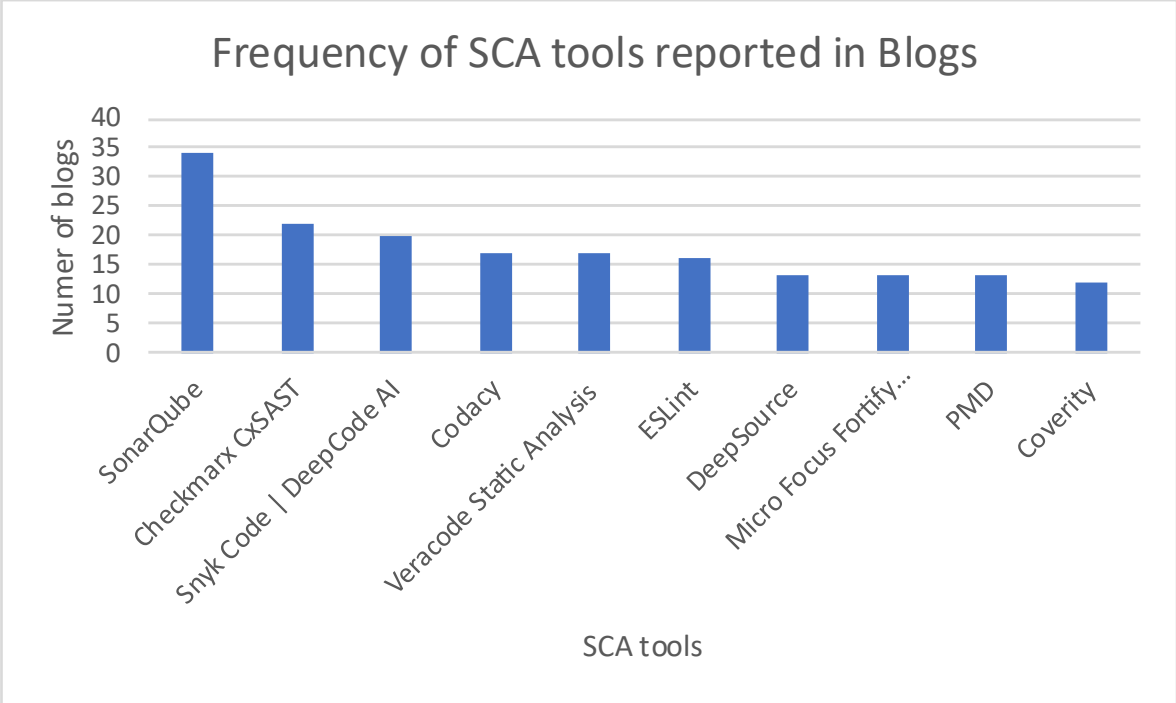

a. Popularity of SCA tools in academia b. Popularity of SCA tools in blogs

Figure 2 Popularity of SCA tools in academia and industry (blog study)

## 3.3 Source code selection

PMD and SonarQube SCA tools were selected for our study, and two publicly available source code datasets were chosen for analysis with the selected SCA tools. Having two separate code datasets will enable us to ascertain the generalisability of our outcomes. The first dataset is the Stack Overflow dataset collected by Meldrum et al. [2], and the second dataset is the Apache Tomcat source code downloaded from GitHub on 8th December 2023, referred to as the Apache Tomcat dataset. These two code datasets were used in our study as they were widely used in academia to evaluate software source codes, and these datasets contain two unique sets of Java codes [47-55], providing evidence for variable code lengths and potential quality. Stack overflow dataset contains source code snippets while Apache Tomcat contains full-length source code.

The first dataset comprises 8,010 Java files collected from code snippets found in posts on the Stack Overflow website[12] which is a question-and-answer website for software engineering, utilised by developers of varying experience levels. In this source code dataset, the number of lines in a file ranges from 1 to 355. Furthermore, five Java files contained only class definitions with no code inside them. Irrespective, those Java files were also utilised for the analysis.

The code datasets should be compiled to scan datasets with SonarQube. However, there were some compilation issues with the Stack Overflow code dataset. In this code dataset, there were some files with conflicting file names. For example, there were two “Employee” classes in two files. In such scenarios, the conflicting class name was modified by adding a number after the class name (“Employee1”). Additionally, when the class was modified with the “public” modifier, but the file name did not match the class name, the “public” modifier was removed. We did not modify the file name as we needed to keep references to the original code dataset. Moreover, we added some library import statements where necessary.

The second dataset is the Apache Tomcat source code downloaded from GitHub on the 8th December 2023[13]. The code dataset comprises 2,749 Java files, all of which are compilable.

[12] https://stackoverflow.com/

[13] https://github.com/apache/tomcat

The number of lines in each file ranges from 20 to 6,285, including comment lines. Furthermore, only the Java files were utilised for our study.

It is reported that the code snippets found in Stack Overflow are of low quality [2]. In contrast, the Apache Tomcat dataset is developed by a software engineering community that reviews the source code regularly[14]. Therefore, it should be of higher quality. Consequently, we have selected two datasets at different maturity levels.

The objective of this study is to create warning datasets generated by PMD and SonarQube. However, the actual datasets used is not crucial, as our primary aim is to capture variations in code length and style in detecting false-positive warnings from best-performing tools to explore rules of best tools and their compliance to established coding standards. Accordingly, we used source code datasets that are widely used in academia, which enriches existing knowledge in the space [47-55].

### 3.4 Identify standards for individual rules

Existing studies have used several approaches to annotate SCA warnings [13]. According to Lin et al. [13], existing studies have used automatic labelling frameworks and manual labelling methodology to create warning datasets, where automatic labelling was seen to have a higher probability of incorrect labelling than manual labelling. Thus, manual labelling was done by the researchers, albeit they did not use any standards/guidelines [28, 33]. Therefore, to annotate the datasets in the current study, the standards listed in Table 2 were used. Using established standards ensures that the labels are consistent and adhere to existing standards.

In identifying relevant standards, we referred to the official documentation for each rule in PMD and SonarQube to determine whether the authors used any standards to define their rules, where some standards were identified from the documentation. OWASP was added to the list because it was used in existing studies [56, 57], and the Java official documentation is also used as a standard. Finally, there were some papers (ID: 10 and 11 in Table 2) mentioned in some PMD rules as a further reference. Since those papers were acceptable studies, those standards were also selected.

These standards were used to validate the SCA tool's rules, as these standards have been widely adopted by the software development industry and are considered reliable [56-58]. However, SonarQube and PMD introduce new rules based on their experience, and they also modify rules based on community feedback[15,16]. These rules are typically not yet mature enough to be presented as standards, and in fact some rules contradict others[17]. For example, the "AtLeastOneConstructor" rule in PMD says that "Each non-static class should declare at least one constructor. Classes with solely static members are ignored, refer to UseUtilityClassRule to detect those." However, this rule contradicts the rule "UnnecessaryConstructor" in PMD. Therefore, rules in SonarQube and PMD cannot be considered standards. Consequently, we evaluate generated warnings for our datasets against more established standards.

Table 2 Established Java Coding Standards

| ID | Established Standards | Reason for selecting the standard |
|---|---|---|

[14] https://github.com/apache/tomcat
[15] https://community.sonarsource.com/
[16] https://github.com/pmd/pmd/pulls
[17] https://pmd.github.io/pmd/pmd_rules_java.html

| | | |
|---|---|---|
| 01 | https://wiki.sei.cmu.edu | Wiki page maintained by Carnegie Mellon University and used as a standard for SonarQube |
| 02 | Book: Clean Code | Used as a standard for SonarQube |
| 03 | Book: Effective Java by Joshua Bloch | Used as a standard for SonarQube |
| 04 | Book: Mastering Regular Expressions by Jeffrey Friedl | Used as a standard for SonarQube |
| 05 | Book: Refactoring: Improving the Design of Existing Code by Martin Fowler, Kent Beck, John Brant, William Opdyke, and Don Roberts. | Used as a standard for SonarQube |
| 06 | Google Java Coding Style Guide | Used as a standard for SonarQube and A popular and well-accepted standard [59, 60] |
| 07 | https://cwe.mitre.org/data/definitions/ | Used as a standard for SonarQube and A popular and well-accepted standard [61, 62] |
| 08 | Java Doc | Java official user guide |
| 09 | Book: Object-Oriented Metrics in Practice: Using Software Metrics to Characterize, Evaluate, and Improve the Design of Object-Oriented Systems by Michele Lanza and Radu Marinescu. | Used as a reference for PMD |
| 10 | Paper: Linguistic Antipatterns - What They Are and How Developers Perceive Them | [63] |
| 11 | Principle of Least Astonishment | [64] |
| 12 | SOLID principles | Used as a standard for SonarQube |
| 13 | Transitive property of equality | Used as a standard for SonarQube |
| 14 | https://owasp.org | A popular and well-accepted standard [56, 57] |

According to the standards shown in Table 2, there were rules that can be classified as follows:

1. SCA rules, which align with the standards.
2. SCA rules, which are against the standards.
3. SCA rules, which neither align with nor against the standards.

The approach used to identify false positive warnings is documented in Table 3. As mentioned in Table 3, if the SCA rule was aligned with the existing standards, SCA warnings were checked manually according to the standards, where warnings were determined as true positive or false positive. However, when the SCA rules were against the existing standards, alarms reported under those rules were removed, and the alarms reported under those rules were considered false positives. Those rules are mentioned in Table 4. Additionally, some SCA rules did not align with or contradict the standards. SCA warnings reported under those rules were checked manually according to the rules' description and the default parameters mentioned on the official website of the SCA tools. Additionally, identifying or defining standards for those rules is an open research area, as those rules are created based on the practices followed by the software engineering Practitioner Community.

Table 3 Approaches used to determine true positives or false positives

| **SCA rule type** | **Approach** |
|---|---|
| SCA rules, which are aligned with the standards. | Code violations were checked manually, according to the standards. |
| SCA rules, which are against the standards. | Removed the code violations as they are against the standards. |
| SCA rules, which are neither aligned with nor against the standards. | According to the SCA tool's documentation and default parameters on the official website, code violations were checked manually. |

Additionally, Deprecated ules, Rule Templates, and Beta rules were removed from the SonarQube rules list, as the SonarQube development community does not recommend using them[18,19]. Those rules are reported in Table 4. When analysing the Java code with PMD, the

[18] https://community.sonarsource.com/t/how-to-fix-deprecated-rules/28871
[19] https://docs.sonarsource.com/sonarqube/latest/user-guide/rules/overview/

"rulesets/internal/all-java.xml" ruleset was used, which contains all non-deprecated rules. Therefore, the report had no warning suggestions related to deprecated rules.

Table 4 Reason for rule removal

| SCA tool | Rule with description | Reason for removal from further analysis |
|---|---|---|
| PMD | AtLeastOneConstructor - Each non-static class should declare at least one constructor. Classes with solely static members are ignored; refer to UseUtilityClassRule to detect those. | According to the official Java Development Kit (JDK) documentation, it is not recommended to create a default constructor as the JDK generates it. |
| | MissingSerialVersionUID - Serializable classes should provide a serialVersionUID field. The serialVersionUID field is also needed for abstract base classes. Each class in the inheritance chain needs its own serialVersionUID field. See also the rule – “Should an abstract class have a serialVersionUID”. | According to the official documentation of the Java Development Kit (JDK), manually defining the serial version ID is not recommended, as the JDK generates it automatically. |
| SonarQube | An open curly brace should be located at the beginning of a line. | The rule does not follow the Google Java Coding Style Guide. |
| | Close curly brace and the following "else", "catch", and "finally" keywords should be on two different lines. | The rule does not follow the Google Java Coding Style Guide. |
| | Custom resources should be closed. | Rule Template |
| | Track comments matching a regular expression. | Rule Template |
| | Track uses of disallowed classes. | Rule Template |
| | Track uses of disallowed constructors. | Rule Template |
| | Track uses of disallowed methods. | Rule Template |
| | @EnableAutoConfiguration should be fine-tuned. | Deprecated |
| | Abstract classes without fields should be converted to interfaces. | Deprecated |
| | Classes should not be loaded dynamically. | Deprecated |
| | finalize should not set fields to "null". | Deprecated |
| | Primitives should not be boxed just for "String" conversion. | Deprecated |
| | super.finalize() should be called at the end of "Object.finalize()" implementations. | Deprecated |
| | XML parsers should not load external schemas. | Deprecated |
| | Classes should not depend on an excessive number of classes (aka Monster Class). | Beta |
| | Methods should not perform too many tasks (aka Brain method). | Beta |
| | The Singleton design pattern should be used with care. | Beta |

## 3.5 Statistical analysis for estimating population false positive percentage

Warnings were manually labelled and the false-positive (FP) percentage for each dataset was calculated, multiplying the number of false positives by 100 and dividing the result by the sample size. To estimate population-level uncertainty from the observed samples, 95% Ionfidence Intervals (CIs) for each FP percentage value were computed using the Wilson score interval method, which provides more accurate bounds for binomial proportions than the normal approximation [65]. The CI for a proportion *p* with sample size *n* was calculated as:

$$CI = \frac{p + \frac{z^2}{2n} \pm z\sqrt{\frac{p(1-p)}{p} + \frac{z^2}{4n^2}}}{1 + \frac{z^2}{n}}$$

We used 1.96 for z, the typical value recommended [65], and the corresponding 95% confidence level was adopted to sample datasets. These intervals quantify the uncertainty of the sample estimates and allow extrapolation to the population of warnings [65]. This statistical analysis was performed in Microsoft Excel using the equation mentioned above. The resulting confidence intervals are reported in Section 3.6.

## 3.6 Sampling and building the dataset

It is not practical to manually check the entire dataset for false positives. Therefore, each dataset was sampled randomly with 95% confidence level and 5% error rate. The **sample size** was calculated statistically using Cochran's formula and divided into each rule in proportion to the number of code violations per each rule [66-68]. When the calculated sample size for a given rule resulted in a fractional value, the number was rounded up to the nearest whole number to ensure each rule was represented in the sample. Within each rule category, random selection was implemented in Microsoft Excel by generating random numbers for each entry using the RAND() function, sorting the data based on these values, and selecting the top entries corresponding to the allocated sample size for each rule. The number of samples per rule for all datasets is provided in our replication package[20] [17].

The datasets were analysed using PMD version 6.55.0 and SonarQube version 10.2.1, downloaded in December 2023. These were the latest stable versions at the time of the investigation and were used to analyse the source code discussed in Section 3.3. Although newer versions of these tools have been developed, the rules studied in this paper remain relevant at the time of manuscript submission. Two SCA tools were used to analyse two source code datasets, producing four SCA warning datasets. A summary of the procedure followed to create datasets is shown in Figure 3.

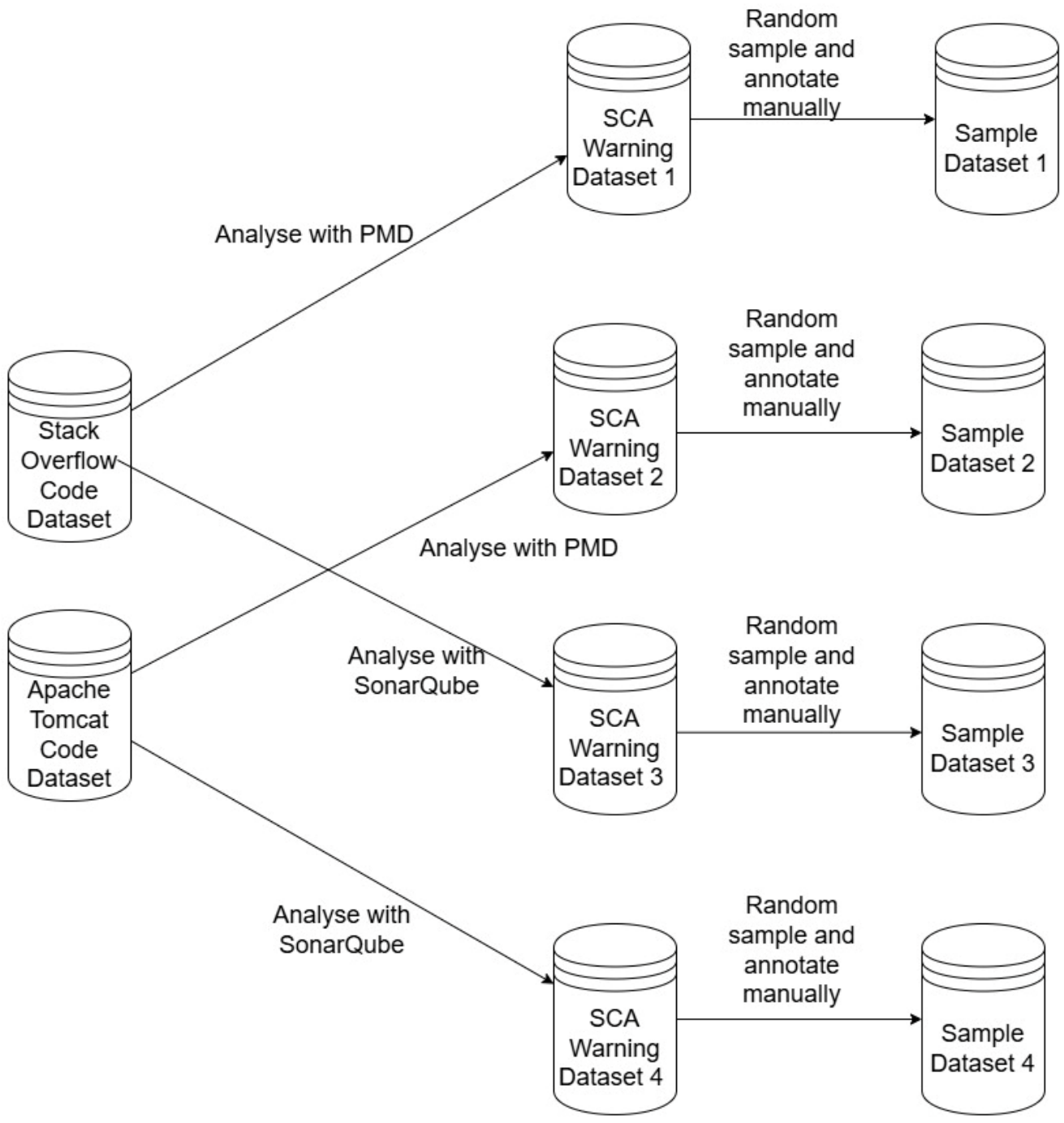

Figure 3 Procedure of the dataset building

For the **Stack Overflow dataset**, 85,584 code violations were found with PMD using the "rulesets/internal/all-java.xml" ruleset, which contained all non-deprecated rules. Those code violations were found for 190 rules under seven rule sets - Code Style, Documentation, Design, Best Practices, Performance, Error Prone, and Multithreading. As discussed in Section 3.4, two of those rules were removed because they were against established standards. After the code violations under these rules were removed, there were 80,788 total code violations. However, manually annotating the entire dataset is not practical. The practical solution is to randomly create a sample with a 95% confidence level and a 5% error margin, as described above, which was done resulting in 531 samples being randomly selected.

[20] Replication package >> Appendices >> Appendix G_Sample sizes for PMD.docx and Appendix H_Sample sizes for SonarQube.docx

Table 5 Wilson score interval for false positive percentages for samples

| Dataset | FP Percentage | Sample size | Confident Interval |
|---|---|---|---|
| 1 | 18.45 | 531 | [15.39, 21.98] |
| 2 | 13.43 | 536 | [10.8, 16.58] |
| 3 | 4.64 | 646 | [3.27, 6.55] |
| 4 | 6.97 | 631 | [5.24, 9.23] |

A given SCA warning was labelled as a true positive if the warning description follows the defined standards. Additionally, in case there are no defined standards for the related SCA rule, then the SCA rule description is used instead of defined standards to label SCA warnings as true positives. Otherwise, SCA warnings were labelled as false positives. These samples were labelled manually with “1” for false positives and “0” for true positives using the approaches mentioned in Table 3. Of the total selected for manual analysis, 98 were false positives. Therefore, the false positive rate in the sample dataset is 18.45%, and there should be a similar false positive rate in the population dataset with a 95% confidence level. The Wilson score intervals for Dataset 1 are calculated as described in Section 3.5 and the results are shown in Table 5. Here it is shown that the false positive percentage in the population Dataset 1 should be between 15.39% and 21.98%.

Following the same procedure, the same dataset was analysed using SonarQube for all rules except Deprecated rules, Beta rules and Rule Templates, and the total number of violations was 108,944. Those violations were found under 312 rules. However, two rules, which were against the established standards, were removed later, as discussed in Section 3.4. Therefore, the total number of code violations was 83,890. Like Dataset 1, a sample with 646 violations was randomly selected with a 95% confidence level and a 5% error margin. The sample dataset was manually checked for false positives. Of that, 30 samples were false positives, and the false positive rate was 4.64%. Thus, there should be a similar false positive rate in the population dataset with a 95% confidence level. Moreover, according to Table 5, the false positive percentage in the population Dataset 2 should be between 10.8% and 16.58%.

Similarly, the **Apache Tomcat dataset** was analysed using PMD and SonarQube with the same set of rules. The total number of code violations was 177,506 and 221,954, respectively, which fell under 199 and 283 rules in PMD and SonarQube. Two rules which are against the established standards for each SCA tool were removed in a similar manner to the previous dataset, and there were a total number of code violations of 176,293 and 128,524, respectively. Much like the previous datasets, samples were selected randomly, resulting in 536 and 631 for PMD and SonarQube, respectively. After manually checking the code violations, there were 72 false positives (13.43%) for PMD and 44 false positives (6.97%) for SonarQube for the Apache Tomcat dataset. Therefore, the population datasets of the sample datasets should have the same false positive rate with a 95% confidence level and 5% error rate. Furthermore, according to Table 5, the false-positive percentage in population Dataset 3 should be between 3.27% and 6.55% and in population Dataset 4, it should be between 5.24% and 9.23%. A simple interface was developed to support the manual annotation process. It shows all the relevant information on one page. Then we provided a suitable label in the software, and the label was recorded with the required details, such as the issue number and warning message. Using this software reduced mistakes that can occur during manual annotations.

A **summary of the four datasets** is presented in Table 6 and Table 7. This summary identifies the sample datasets: Dataset 1 consists of warnings produced by PMD after analysing the Stack Overflow dataset, while Dataset 2 comprises warnings generated by PMD for Apache Tomcat. Meanwhile, Dataset 3 and Dataset 4 represent the warnings created by SonarQube for the Stack Overflow and Apache Tomcat datasets, respectively. These sample datasets were deeply analysed to generate results discussed in the Section 4 and were used to answer research questions outlined in the Section 3.1. In addition, Table 8 provides a comprehensive overview of the variables used to answer each research question in this study.

Table 6 Summary of datasets

| Dataset Name | SCA tool | Dataset | Total alarm count | # of rules found |
|---|---|---|---|---|
| Dataset 1 | PMD | Stack Overflow dataset | 85,584 | 190 |
| Dataset 2 | PMD | Apache Tomcat | 177,506 | 199 |
| Dataset 3 | SonarQube | Stack Overflow dataset | 108,944 | 312 |
| Dataset 4 | SonarQube | Apache Tomcat | 221,954 | 283 |

Table 7 Summary of sample datasets

| Dataset Name | # of rules removed (except Beta, Deprecated, and Rule Template rules) | # of alarms after rule removal | Sample size | Number of false positives in the sample | False positive percentage |
|---|---|---|---|---|---|
| Dataset 1 | 2 | 80,788 | 531 | 98 | 18.45 |
| Dataset 2 | 2 | 176,293 | 536 | 72 | 13.43 |
| Dataset 3 | 2 | 83,890 | 646 | 30 | 4.64 |
| Dataset 4 | 2 | 128,524 | 631 | 44 | 6.97 |

Table 8 Measures for answering research questions

| Metric | Relevant research question |
|---|---|
| SCA tool | RQ1, RQ2, RQ3, RQ4 |
| SCA rule | RQ1, RQ2, RQ3, RQ4 |
| Number of usable rules | RQ1 |
| Number of rules which generated code violations | RQ1, RQ2 |
| Rule standards | RQ1, RQ2, RQ4 |
| Are established standards aligned with or against established standards | RQ1, RQ2 |
| Number of rules per standard | RQ1 |
| Rule category | RQ1 |
| Number of rules per rule category | RQ1 |
| Sample size | RQ2 |
| Number of false positives in the sample | RQ2, RQ3 |
| Rules which generated false positives | RQ2, RQ3 |
| Reason for generating the false positive (for each data point) | RQ3 |
| Gaps , if any, with the SCA rule and established standard | RQ4 |

# 4 Results

Four sample datasets were created, as explained in Section 2, and the datasets were analysed to answer the research questions outlines in Section 3.1 as follows.

## 4.1 RQ1: What are the rules and standards of SCA tools that generate warnings for specific source code datasets?

For PMD, we used the "rulesets/internal/all-java.xml" rulesets to identify all code quality violations in both code bases and example rules in PMD and code examples with related issues are shown in Table 9. PMD checks the source code against rules defined, and the related issue line is identified. PMD then return the issue with issue line, rule, and a description. There were 271 total rules in PMD, and out of those rules, 221 rules generated code violations. Of the 221 rules, 178 comply with at least one standard referred to in Section 3.4. Furthermore, two rules violated those standards, and 41 rules neither violated nor adhered to them.

Table 9 Examples for PMD rules and source codes with related issues – AddEmptyString, ArrayIsStoredDirectly, AssignmentInOperand, CommentRequired

| **PMD rule** | **Example source code** |
|---|---|
| AddEmptyString | public class C355962{<br>String result = (System.currentTimeMillis() * 1000) + ""; // issue_line<br>} |
| ArrayIsStoredDirectly | public class FooA {<br>    public int[] mData;<br>    public FooA(int[] arr) {<br>        System.out.println("aa");<br>        mData = arr;// issue_line<br>    }<br>} |
| AssignmentInOperand | public class C250127{<br>public boolean isPalindrome(String s) {<br>    s = s.replaceAll("[^a-zA-Z0-9]", "").toLowerCase(); // issue_line<br>    int b = 0;<br>    int e = s.length()-1;<br><br>    while(b < e) {<br>        if(s.charAt(b++) != s.charAt(e--))<br>            return false;<br>    }<br>    return true;<br>}<br><br>} |
| CommentRequired | public class C299113{<br>String hot89_9= new String("Hot 89.9"); //issue_line<br>} |

For SonarQube, there were 632 rules for the Java language. Of that sum, 359 rules generated code violations for the two source code datasets, where 290 rules adhered to the standards outlined in Section 3.4. Moreover, two rules violated those standards, and 67 rules neither violated nor adhered to the standards. Furthermore, deprecated rules, Beta rules and Rule templates were excluded before analysis, and there were 15 rules belonging to these categories. Additionally, Table 10 shows some example rules in SonarQube along with related code examples and related issue lines, with the outcome of SonarQube being similar to PMD.

Table 10 Examples for SoanrQube rules and source codes with related issues – Method names should comply with a naming convention, Standard outputs should not be used directly to log

anything, An open curly brace should be located at the end of a line, Wildcard imports should not be used

| SonarQube rule | Example source code |
|---|---|
| Method names should comply with a naming convention | public class C102762{<br>public boolean AnswerCall(String r) { // issue_line<br>return r.equals("Brother");<br>}<br>} |
| Standard outputs should not be used directly to log anything | public class C36028{<br>static void data() {<br>int array[] = {1,5,6};<br>int alength = array.length;<br>System.out.println("  Location\tData"); // issue_line<br>for(int i=0;i<alength;i++) {<br>System.out.println("  " + i + "\t\t" + array[i]);<br>}<br>}<br><br>} |
| An open curly brace should be located at the end of a line | public class C101875{<br>public void remove()<br>{ //Issue_line<br>throw new UnsupportedOperationException();<br>}<br>} |
| Wildcard imports should not be used | import java.awt.image.BufferedImage;<br>import javax.swing.*; // issue_line<br>import javax.swing.border.*;<br><br>public class ThreePartBorder { |

The complete set of rules and their compliance with standards are provided in our replication package[21] [17], and a summary of the outcomes is provided in Table 11, where it is shown that SonarQube has more rules than PMD. Additionally, PMD identified code violations across nearly all its rules for the code datasets, while only around 50% of SonarQube rules highlighted violations in the same two datasets.

Table 11 Summary of rules and compliance in PMD and SonarQube

| SCA tool | Number of usable rules | Number of rules which generated code violations | Number of rules with standards | Number of rules without standards | Number of rules against standards |
|---|---|---|---|---|---|
| PMD | 271 | 221 | 178 | 41 | 2 |
| SonarQube | 632 | 359 | 290 | 67 | 2 |

There were three types of rules, as mentioned in Table 3. Those types were: (1) SCA rules which are aligned with the standards, (2) SCA rules which are against the standards, and (3) SCA rules which are neither aligned with nor against the standards. Following these rule types, the number of rules with standards and the number of rules without and against standards are mentioned in Table 11. The established standards, used to validate rules, are discussed in Section 3.4. In our study, only the SCA rules that generated warnings for our two code datasets were considered for identifying standards. Among these rules, the majority had at least one defined standard. However, a significant number of rules do not have defined standards. Additionally, only two rules for each SCA tool contradicted the established standards.

Appendix A and Appendix B demonstrate the established standards and the number of rules related to the standards for PMD and SonarQube, respectively. Furthermore, the number of rules is calculated such that any rule associated with multiple standards is counted multiple times, once per standard. According to Appendix A, there were 12 standards used for PMD. Furthermore, as shown in Appendix B, a total of 11 standards were utilised for SonarQube. A comparison of Appendix A and Appendix B is shown in Figure 4. Figure 4 demonstrates the top 10 standards used to annotate code violations generated by PMD and SonarQube rules. According to Figure 4, the most used standard was JDK Documentation, and the second was the SEI website[22] for both PMD and SonarQube. Moreover, OWASP[23] has the lowest rule count. In addition, SonarQube has more rules than PMD in total, and consequently, SonarQube has the highest rule counts under most standards.

[21] Replication package >> PMD and SonarQube rules with standards.xlsx

We further discussed in detail the number of rules that produced warnings and their distribution across the rule types below, which were classified based on the applicability of the established standards. The rules of SCA tools were categorised by the vendors of the SCA tools based on properties of these rules. These categories, along with the number of rules belonging to these categories, are mentioned in Table 6 for PMD and Table 7 for SonarQube.

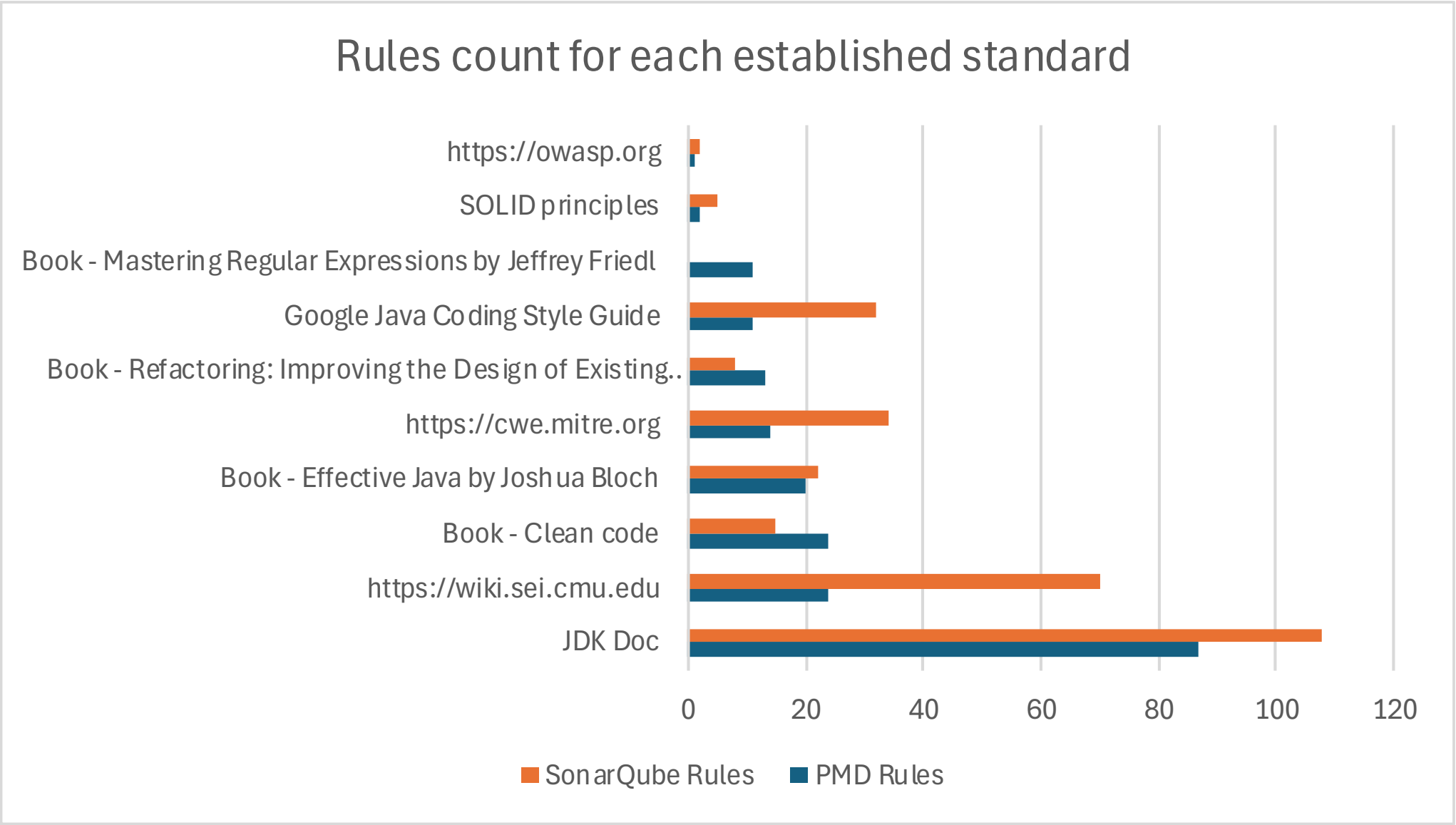

Figure 4 Top 10 standards used to classify rules

Table 12 Different rule categories of PMD

| Rule type | Number of Rules generating warnings |
|---|---|
| Best Practices | 43 |
| Code Style | 46 |
| Design | 39 |
| Documentation | 4 |
| Error Prone | 57 |
| Multithreading | 10 |
| Performance | 22 |

[22] https://wiki.sei.cmu.edu
[23] https://owasp.org

According to the official website of PMD[24], there are seven categories of rules in PMD, and those rules, along with the number of rules which reported warnings for our two datasets, are shown in Table 12. Rules that reported warnings for our two datasets covered all the types of rules in PMD. In addition, most of these rules belong to the "Error Prone" category. Moreover, the "Documentation" category has only four rules.

According to the SonarQube official website[25], there are four types of rules in SonarQube: "Code Smell" (maintainability domain), "Bug" (reliability domain), "Vulnerability" (security domain), and "Security Hotspot" (security domain). According to Table 13, rules that reported warnings for our two datasets covered the "Code Smell", "Bug", and "Vulnerability" aspects of the source code quality for SonarQube. A total of 276 rules, representing the majority of warnings in our two source code datasets belong to the "Code Smell" category, and 65 and four rules belong to the "Bug" and "Vulnerability" categories, respectively. Even though there were no warnings in the "Security Hotspot" category, our warning datasets consist of warnings from most categories. In comparison to existing studies, building ML models to identify false positive warnings utilised datasets with a limited category of SCA warnings. However, our datasets include SCA warnings corresponding to most of the categories in Table 7. According to the existing studies summarised in Section 2, most of the warning datasets used to train ML models to identify false positives lack diversity in rules, as the datasets consist of warnings from only a few rules. Consequently, our four datasets would support the training of ML models for classifying and identifying false positive warnings.

Table 13 Different rule categories of SonarQube

| **Rule type** | **Number of Rules generated warnings** |
|---|---|
| Bug | 65 |
| Code Smell | 276 |
| Vulnerability | 4 |
| Security Hotspot | 0 |

## 4.2 RQ2: Which rules are prone to false positives?

Previous studies have shown that SCA tools generate many false-positive SCA warnings [3-5]. However, we identified that not all the rules generate false suggestions. In both PMD and SonarQube, specific rules generate false positives, as shown in Appendix C (for PMD) and Appendix D (for SonarQube). In both Appendix C and Appendix D, the term "sample count" refers to the number of samples calculated for each rule in Section 3.6. Furthermore, these appendices show that these rules generate false positives only sometimes, and the percentage of false positives is reported in the False Positive (FP) column, which is the ratio of false positives to samples for each rule. Additionally, the top 10 rules which are prone to false positives are shown in Figure 5 for PMD and Figure 6 for SonarQube.

Figure 5 presents the false positive warning distribution for Dataset 1 and Dataset 2, which are the sample datasets generated by analysing the Stack Overflow code dataset and the Apache Tomcat code dataset using PMD, as mentioned in the Table 6. In Figure 5, the vertical axis represents the number of warnings that generated false positives, while the percentage of false positives for each rule, relative to all identified false positives, is shown on top of the corresponding bars. Figure 5 shows that for both Stack Overflow (Dataset 1) and Apache

[24] https://pmd.github.io/pmd/pmd_rules_java.html
[25] https://docs.sonarsource.com/sonarqube-server/10.2/user-guide/rules/overview/

Tomcat (Dataset 2) code, the rule with the highest number of reported false positives was “CommentRequired”, followed by “LawOfDemeter”. However, only the top two rules are common to both datasets. The rule that generated the third highest number of false warnings for Stack Overflow code is “UseUtilityClass”, while for Apache Tomcat code it is “JUnitTestsShouldIncludeAssert”. The other rules exhibit distinct patterns. Further details are demonstrated in Appendix C.

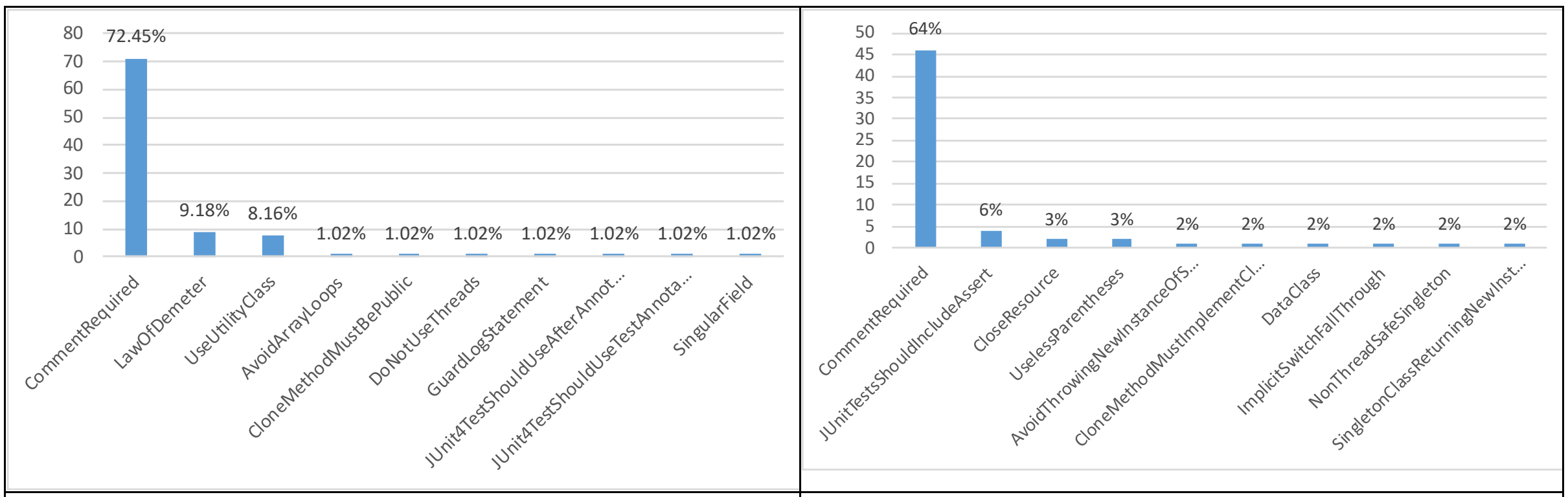

a. PMD false positives for Stack Overflow code

b. PMD false positives for Apache Tomcat code

Figure 5 False positive distribution for PMD datasets (Y-axis shows the false positive count. The percentage values of false positives for each rule, relative to all identified false positives, are shown on top of the corresponding bars.)

Figure 6 visualises the false positive warning distribution for Stack Overflow and Apache Tomcat code for the SonarQube analysis (refer to Table 6). In Figure 6, the vertical axis indicates the number of warnings that generated false positives, while the percentage of false positives for each rule, relative to all false positives identified, is displayed above the corresponding bars. The SonarQube rule “Magic numbers should not be used” generated the highest number of false positives in both datasets. For Stack Overflow code, the second highest false positive is “Source code should be indented consistently”. In contrast, the rule “Track lack of copyright and license headers” is the second most frequently reported cause of false positives in Apache code. Moreover, only five rules produced false positives for Stack Overflow code (Dataset 3), whereas 14 rules generated false positives for Apache Tomcat code (Dataset 4). Additional details are reported in Appendix D.

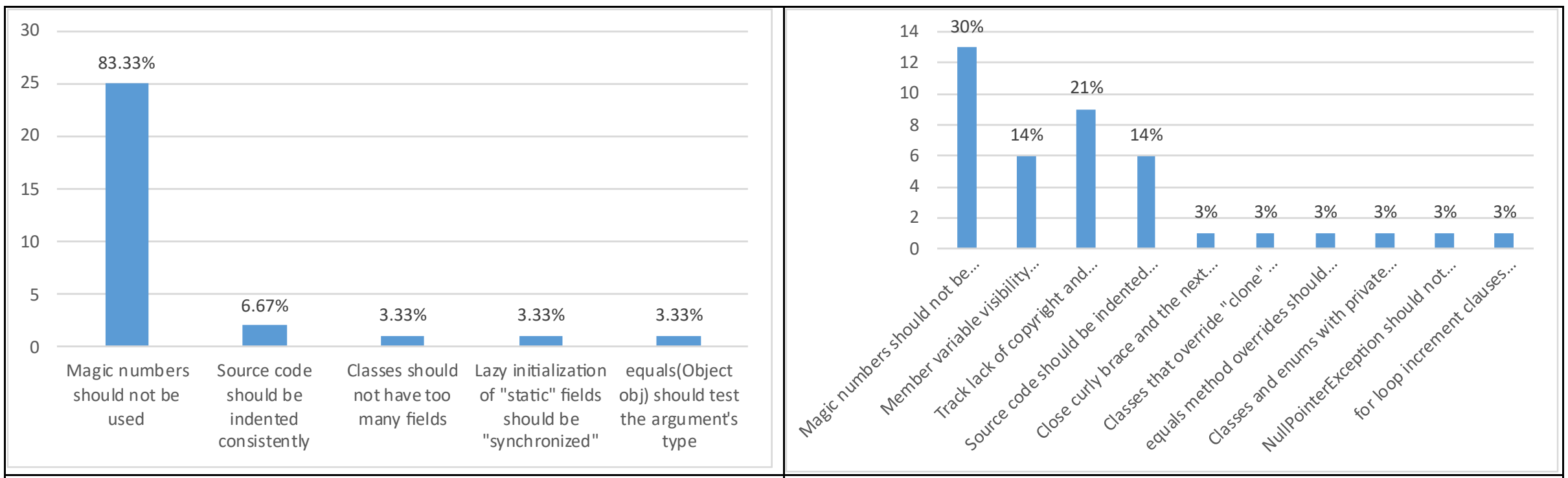

a. SonarQube false positives for Stack Overflow code

b. SonarQube false positives for Apache Tomcat code

Figure 6 False positive distribution for SonarQube datasets (Y-axis shows the false positive count. The percentage values of false positives for each rule, relative to all identified false positives, are shown on top of the corresponding bars.)

Furthermore, as shown in Table 14, PMD produced warnings for 190 rules in Dataset 1 (Stack Overflow code), and a total of 531 samples have been chosen for manual checks to identify false alarms. A total of 98 false positives are seen, caused by 13 rules. Furthermore, some standards have been established for ten rules, but we could not identify any for three rules. A summary of these outcomes and those for SonarQube is shown in Table 14, where the highest false positive rates were reported for PMD, and most of the rules prone to false positives were governed by standards, as demonstrated in Table 14. Other outcomes are seen for SonarQube show that there were fewer false positives for Stack Overflow than Apache Tomcat code. Overall, findings indicate that only a few rules are accountable for generating numerous false positive code violations.

Table 14 Number of rules prone to false positive for rules with/without standards

| **Sample Dataset** | **SCA tool** | **Dataset** | **Number of rules which generated results** | **Sample size** | **Number of false positives** | **Percentage of false positives** | **Number of rules which generate false positives** | **Number of rules with standards** | **Number of rules without standards** | **Percentage of rules that generated false positives** |
|---|---|---|---|---|---|---|---|---|---|---|
| Dataset 1 | PMD | Stack Overflow dataset | 190 | 531 | 98 | 18.45 | 13 | 10 | 3 | 6.84% |
| Dataset 2 | PMD | Apache Tomcat | 199 | 536 | 72 | 13.43 | 16 | 12 | 4 | 8.04% |
| Dataset 3 | Sonar Qube | Stack Overflow dataset | 312 | 646 | 30 | 4.64 | 5 | 3 | 2 | 1.60% |
| Dataset 4 | Sonar Qube | Apache Tomcat | 283 | 631 | 44 | 6.97 | 14 | 12 | 2 | 4.95% |

## 4.3 RQ3: What are the reasons for reported false positives by SCA tools?

A summary of the outcomes to answer this RQ is shown in Table 15 and Table 16, while additional details are provided in our replication package[26] [17]. In Table 15, there are three reasons for false positives generated by PMD. The primary reason for this is that most of the reported SCA warnings do not require attention base on the established standards, allowing these warnings to be safely disregarded without further action. For example, the PMD rule – “CommentRequired” advises adding comments to variables (fields) in the source code. However, it partially aligns with the defined standards, and according to the book “Clean Code” and the JDK documentation, field comments are considered unnecessary, and the variable name should be self-explanatory. Consequently, those alarms have been identified as false positives. However, we have considered the rule in our analysis, because not all the warnings it generates can be ignored under the defined standards. For example, according to established standards, Java main methods require comments, and warnings for missing comments for Java main methods are also reported under the “CommentRequired” rule. The second most common cause of false positive alerts from SCA tools is the tool's misidentification of source code patterns.

[26] Replication package >> Appendices >> Appendix E_Reason for False Positives in PMD.docx and Appendix F_Reason for False Positives in SonarQube.docx

For instance, PMD has misidentified a group of “if” statements in the source code as being deeply nested “if” statements. Although rare, there is another reason SCA tools may generate false alarms (“Rules enforcement conflicts with intended code behaviour”). Some source code implementations might not conform to the SCA tool’s rules, when there is no alternative method to implement this source code correctly. For instance, if a developer needs to copy an array while altering their index positions, it is not feasible to use the “System.arraycopy” Java method as recommended by the PMD rule. Furthermore, this trend is also typical for both Stack Overflow and Apache Tomcat code, as illustrated in Figure 7.

Table 15 Reason for PMD false positives for Stack Overflow and Apache Tomcat code

| **Generalised Reason** | **PMD false positive count in Stack Overflow code** | **PMD false positive count in Apache code** |
|---|---|---|
| It is not required according to the standards | 79 | 51 |
| The tool could not identify code patterns | 18 | 19 |
| Rules enforcement conflicts with intended code behaviour | 1 | 2 |
| **Total** | **98** | **72** |

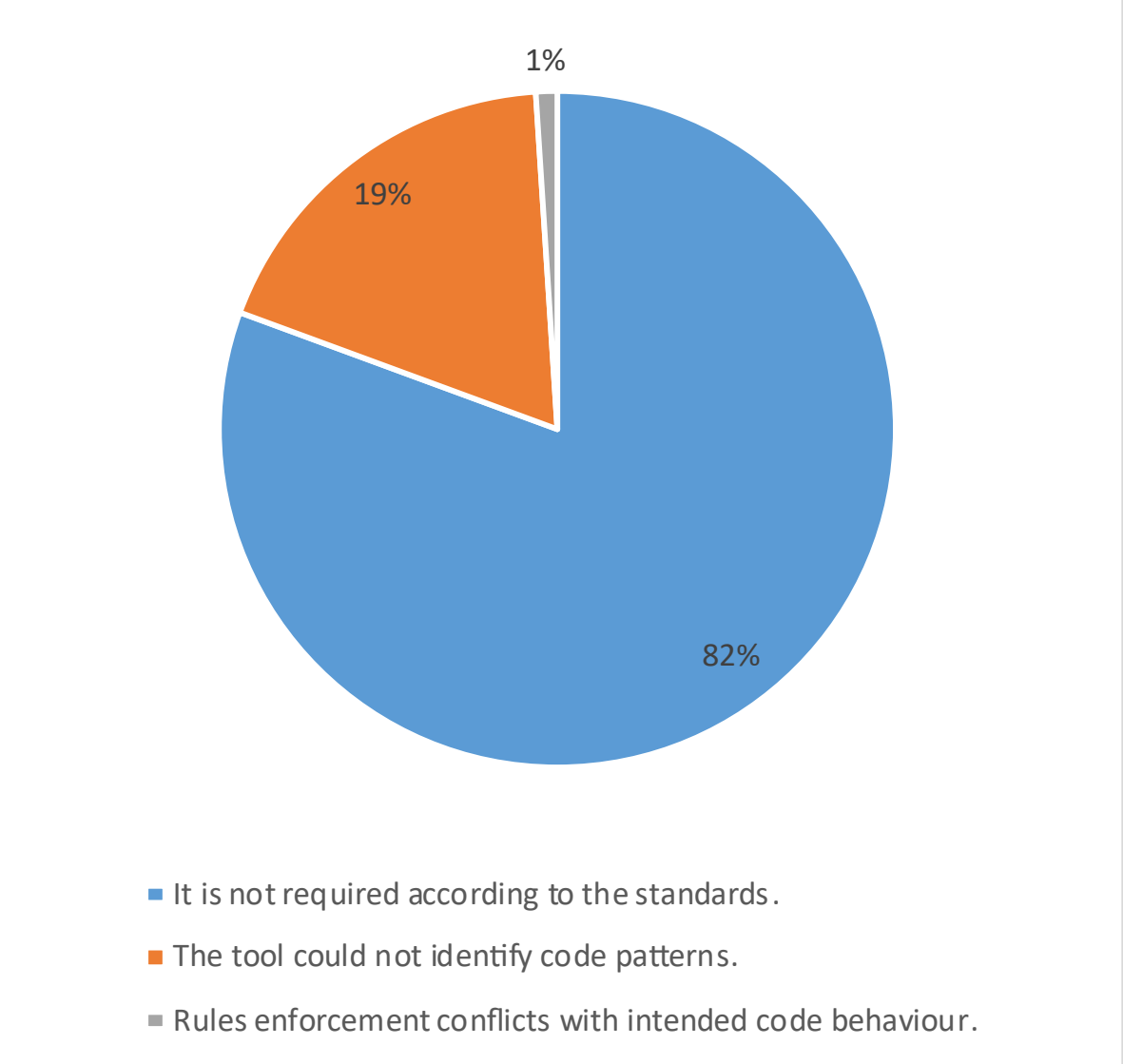

Reasons for false positives for Stack Overflow code

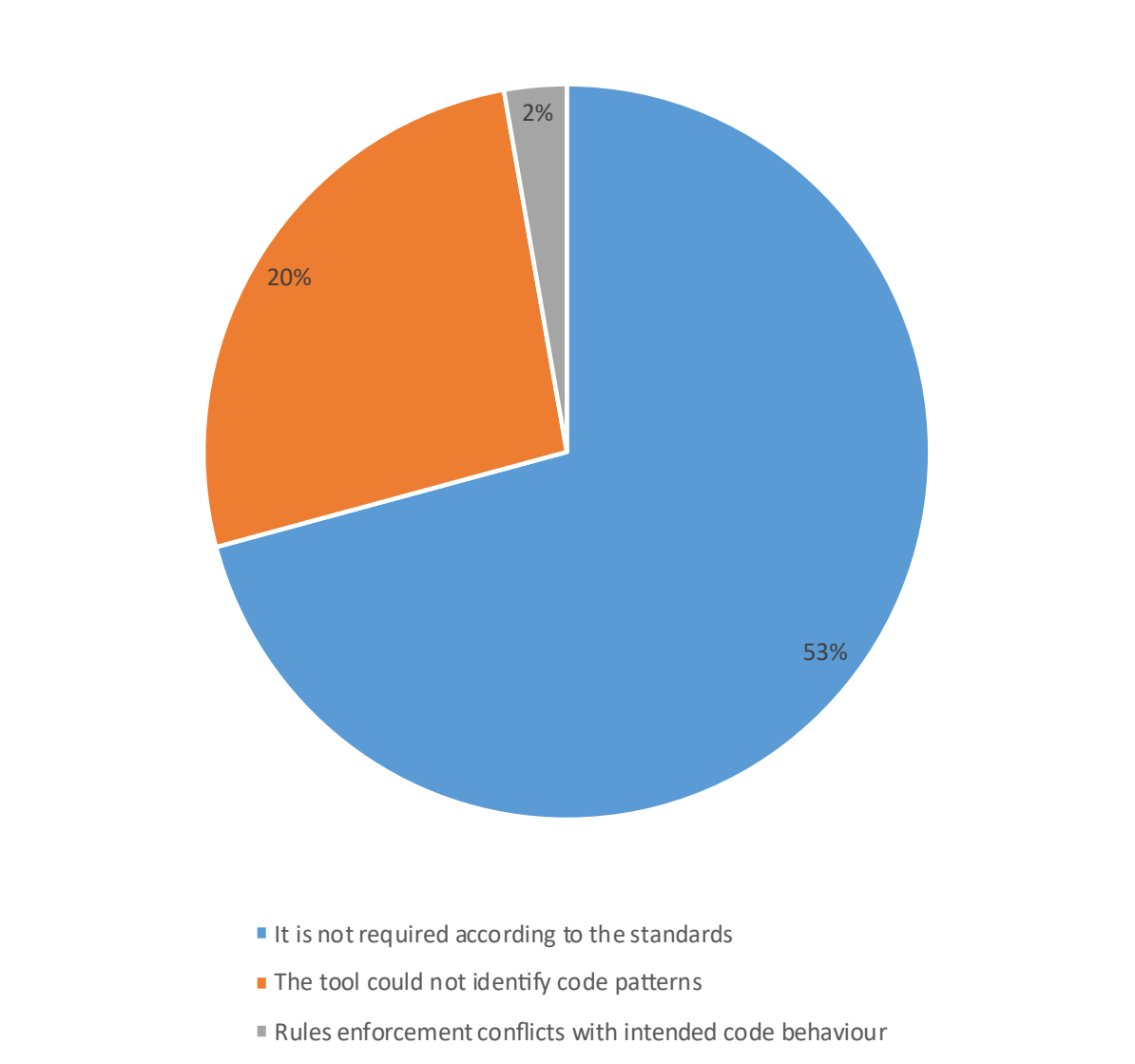

Reasons for false positives for Apache Tomcat code

Figure 7 Reasons for false positives generated by PMD

In Table 16, there are three reasons for SonarQube false positives in Stack Overflow and Apache Tomcat code. The most frequent reason for Stack Overflow code is “Sufficient readability even without following the rule,” whereby even if the rule is not followed, the source code can still be easily understood, despite not adhering to SonarQube's guidelines. As a result, adhering to the rules is unnecessary [69]. Mainly, the SonarQube rule - “Magic numbers should not be used” caused this disagreement. We evaluated the readability of the number reported as a magic number by manually checking whether it was understood by analysing source code near the issue line, such as comments and the name of the element list (if the number belongs to any Java collection). In Apache Tomcat code, SonarQube mostly produced false positives because the SCA tool failed to recognise the code pattern. For instance, SonarQube indicates in some warnings that some code lines are inconsistently indented[27], which is a misleading recommendation. The second most common reason for generating false positive warnings in both datasets is that the source code cannot be implemented by developers in accordance with the SonarQube rule ("Rules enforcement conflicts with intended code behaviour"). If the developer adhered to the rule, the source code would produce unintended results. For example, if the developer needs to copy one Java array to another while modifying data, it is impossible to use the 'Arrays.copyOf' method as recommended by SonarQube. Outcomes here suggest that refining the rules of SCA tools to mitigate these issues could reduce the generation of false warnings by these tools.

[27] https://google.github.io/styleguide/javaguide.html#s4.2-block-indentation

Table 16 Reason for SonarQube false positives of Stack Overflow and Apache Tomcat code

| Generalised Reason | False positive count in Dataset 3 | False positive count in Dataset 4 |
|---|---|---|
| Sufficient readability even without following the rule | 13 | 9 |
| Rules enforcement conflicts with intended code behaviour | 12 | 12 |
| The tool could not identify code patterns | 5 | 23 |
| **Total** | **30** | **44** |

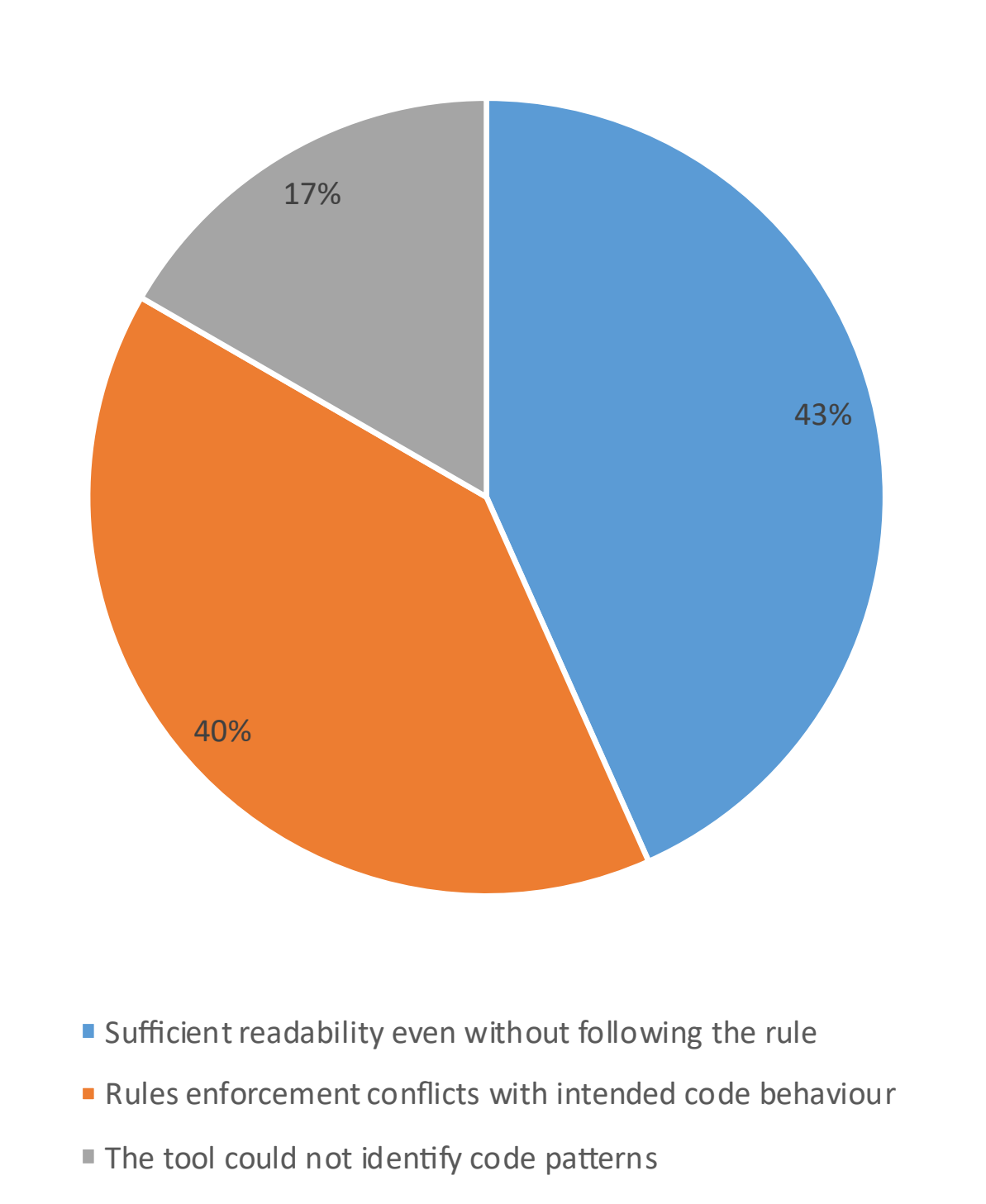

Reasons for false positives for Stack Overflow code

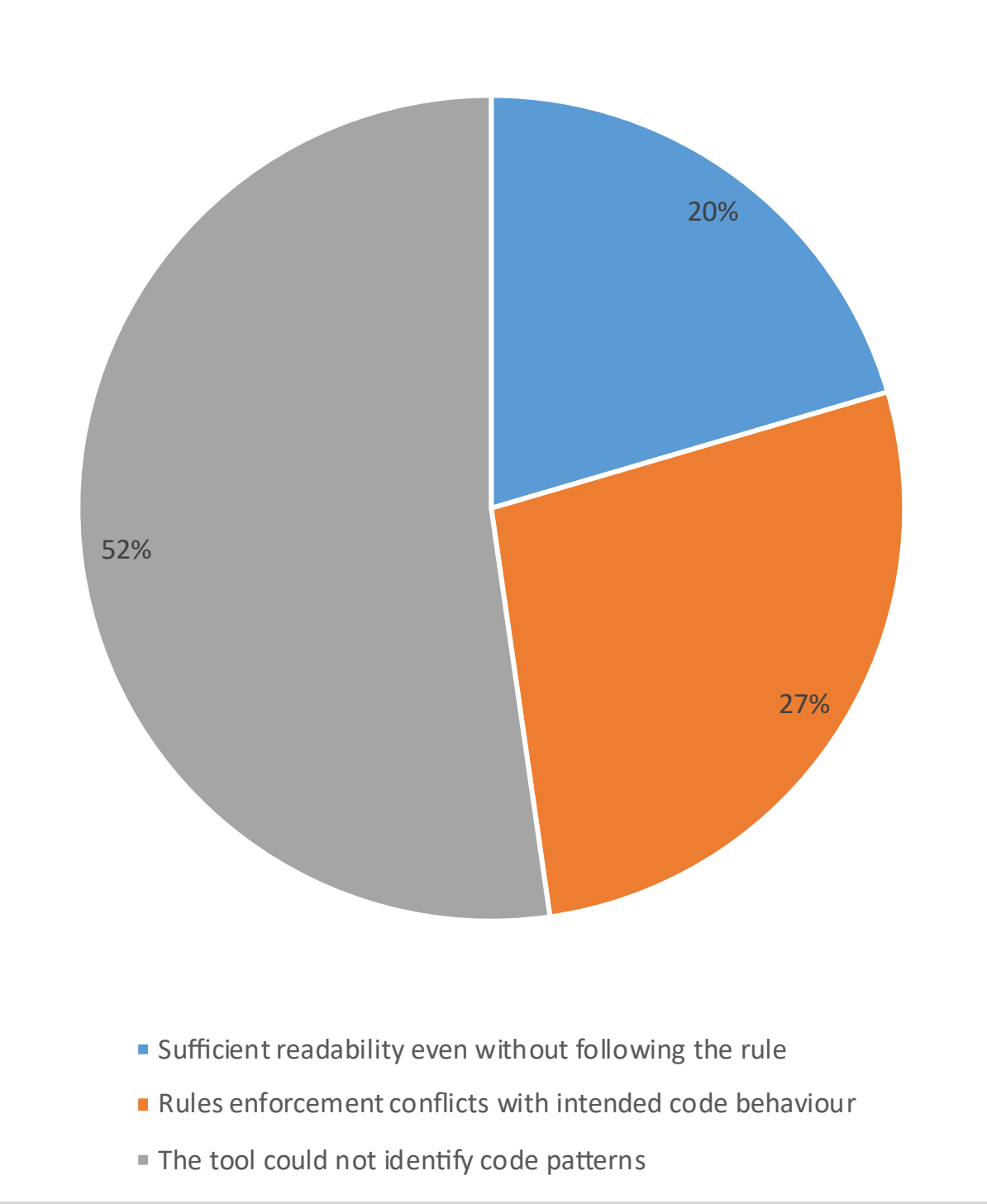

Reasons for false positives for Apache Tomcat code

Figure 8 Reasons for false positives generated by SonarQube

The top five rules shown in Figure 5 and Figure 6 were selected, and the reasons they produced false-positive warnings in PMD and SonarQube are detailed in Table 17 and Table 18, respectively. Here it is shown that the primary reason the majority of rules generate false positives is the tool's inability to identify code patterns in both PMD and SonarQube. Additionally, following some warning recommendations will affect the intended behaviour of the source code. Therefore, this is also a reason why both PMD and SonarQube generate false positives. Moreover, in PMD, many rules generate false positives because those warnings can be ignored when following established standards. Furthermore, in SonarQube, some warnings can be ignored because the source code remains readable even without following them, as expected by the SonarQube rule.

Table 17 Reasons for PMD generating false positives for the top 5 rules in Dataset 1 and Dataset 2

| **Rule** | **Data sets** | **Reason** | **Example source codes** |
|---|---|---|---|
| a. AvoidArrayLoops | Dataset 1 | Rules enforcement conflicts with the intended code behaviour. | public static void main(String[] s)<br>{<br>int[] array = {1,2,3,4,5};<br><br>int temp = array[4];<br><br>for(int i=array.length-1;i>0;i--) // issue_line<br>{<br>array[i]=array[i-1];<br>}<br><br>array[0]= temp;<br><br>for(int i=0;i<array.length;i++)<br>{<br>System.out.println(array[i]);<br>}<br><br>}<br>} |
| b. AvoidThrowingNewInstanceOfSameException | Dataset 2 | The tool could not identify code patterns. | if (getInitParameter(HTTPS_SERVER_PORT_PARAMETER) != null) {<br>try {<br>setHttpsServerPort(Integer.parseInt(getInitParameter(HTTPS_SERVER_PORT_PARAMETER)));<br>} catch (NumberFormatException e) { // issue_line<br>throw new NumberFormatException(sm.getString("remoteIpFilter.invalidNumber",<br>HTTPS_SERVER_PORT_PARAMETER, e.getLocalizedMessage()));<br>}<br>}<br><br>if (getInitParameter(ENABLE_LOOKUPS_PARAMETER) != null) {<br>setEnableLookups(Boolean.parseBoolean(getInitParameter(ENABLE_LOOKUPS_PARAMETER)));<br>}<br>} |
| c. CloneMethodMustBePublic | Dataset 1 | The tool could not identify code patterns. | public class CloneMe {<br>int version;<br>Another another;<br><br>public CloneMe(int newVersion, Another obj) {<br>this.version = newVersion;<br>this.another = obj;<br>}<br>// and so on<br><br>@Override<br>protected Object clone() throws CloneNotSupportedException { // issue_line<br>return super.clone(); // You can also provide your own implementation here<br>}<br>} |
| d. CloseResource | Dataset 2 | Rules enforcement conflicts with the intended code behaviour. | protected void doGet(HttpServletRequest request, HttpServletResponse response)<br>throws ServletException, IOException {<br><br>response.setContentType("text/plain");<br>response.setCharacterEncoding("UTF-8");<br><br>PrintWriter pw = response.getWriter(); // issue_line<br>pw.print("Enter-"); |

| | | | | |
|---|---|---|---|---|
| e. | CommentRequired | Datasets 1 and 2 | It is not required according to the standards. | import java.util.Random;<br><br>public class TheMontyHallParadox {  // issue_line<br>  public static void main(String[] args) {<br>    final int NUMBER_OF_DOORS = 3;<br>    int winBySticking = 0; |
| f. | JUnitTestsShouldInclude Assert | Dataset 2 | The tool could not identify code patterns. | @Test<br>public void testSpecification03() throws Exception {  // issue_line<br>  // Reported as failing during 8.0.11 release vote by Ognjen Blagojevic<br>  // EDH was introduced in 1.0.0<br>testSpecification("EECDH+aRSA+SHA384:EECDH:EDH+aRSA:RC4:!aNULL:!eNULL:!LOW:!3DES:!MD5:!EXP:!PSK:!SRP:!DSS");<br>} |
| g. | LawOfDemeter | Dataset 1 | It is not required according to the standards. | private List<Runnable> drainQueue() {<br>  BlockingQueue<Runnable> q = workQueue;<br>  ArrayList<Runnable> taskList = new ArrayList<>();<br>  q.drainTo(taskList);  // issue_line<br>  if (!q.isEmpty()) {<br>    for (Runnable r : q.toArray(new Runnable[0])) {<br>      if (q.remove(r)) {<br>        taskList.add(r);<br>      }<br>    }<br>  }<br>  return taskList;<br>} |
| | | | The tool could not identify code patterns. | jtf.setForeground(Color.BLUE);<br><br>b.addActionListener(new BoutonListener());<br><br>top.add(label);<br>top.add(jtf);<br>top.add(b);<br><br>this.setContentPane(top);<br>this.setVisible(true);<br>}<br><br>class BoutonListener implements ActionListener {<br>  private final AtomicInteger nbTry = new AtomicInteger(0);<br>  ArrayList<Integer> pins = readPinsData("bdd.txt");<br>  public void actionPerformed(ActionEvent e) {<br>    if (nbTry.get() > 2) {<br>      JOptionPane.showMessageDialog(null, "Number of tries exceeded");<br>      return;<br>    }<br>    final String passEntered=jtf.getText().replaceAll("\u00A0", "");  // issue_line<br>    if (passEntered.length() != 4) {<br>      System.out.println("Pin must be 4 digits");<br>      JOptionPane.showMessageDialog(null, "Ping must be 4 digits");<br>      return;<br>    } |
| h. | UselessParentheses | Dataset 2 | It is not required according to the standards. | boolean verifyErrorIsOptional = (errnum == X509_V_ERR_DEPTH_ZERO_SELF_SIGNED_CERT())  // issue_line |
| i. | UseUtilityClass | Dataset 1 | The tool could not identify code patterns. | import java.util.Map;<br><br>class so1 {  // issue_line |

Table 18 Reasons for SonarQube generating False positives for the top 5 rules in Dataset 3 and Dataset 4

| Rule Description | Dataset | Reason | Example source codes |
|---|---|---|---|
| a. Classes should not have too many fields | Dataset 1 | The tool could not identify code patterns. | static class Gast1 extends JPanel { // issue_line<br><br>// de knoppen, textvelden en labels<br>private JTextField TF1_VNaam, TF1_TussenVoegsel, TF1_ANaam, TF1_Straat, TF1_SNummer, TF1_Postcode,<br>TF1_Plaatsnaam, TF1_GD, TF1_Tel, TF1_Email, TF1_Zoeken;<br>private JLabel L1_VNaam, L1_TussenVoegsel, L1_ANaam, L1_Straat, L1_Postcode, L1_Plaatsnaam, L1_GD, L1_Tel,<br>L1_Email;<br>private JButton B1_Instellingen, B1_Zoeken, B1_Vorige, B1_Volgende;<br><br>public Gast1(final JFrame maker) {<br>super();<br><br>// layout manager uit<br>setLayout(null);<br><br>// Textvelden<br>TF1_VNaam = new JTextField(10);<br>TF1_TussenVoegsel = new JTextField(10);<br>TF1_ANaam = new JTextField(10);<br>TF1_Straat = new JTextField(10);<br>TF1_SNummer = new JTextField(10);<br>TF1_Postcode = new JTextField(10);<br>TF1_Plaatsnaam = new JTextField(10);<br>TF1_GD = new JTextField(10);<br>TF1_Tel = new JTextField(10);<br>TF1_Email = new JTextField(10);<br>TF1_Zoeken = new JTextField(10); |
| b. Close curly brace and the next "else", "catch" and "finally" keywords should be located on the same line | Dataset 2 | The tool could not identify code patterns. | } catch (ProtocolException e) {<br>// Protocol exceptions normally mean the client sent invalid or<br>// incomplete data.<br><br>getLog().debug(sm.getString("abstractConnectionHandler.protocolexception.debug"), e);<br>}<br>// Future developers: if you discover any other<br>// rare-but-nonfatal exceptions, catch them here, and log as<br>// above.<br>catch (OutOfMemoryError oome) { // issue_line<br>// Try and handle this here to give Tomcat a chance to close the<br>// connection and prevent clients waiting until they time out. |
| c. equals(Object obj) should test the argument's type | Dataset 1 | The tool could not identify code patterns. | @Override<br>public boolean equals(Object other) { // issue_line<br>boolean result = (other instanceof UniqueId);<br>if ( result ) {<br>UniqueId uid = (UniqueId)other;<br>if ( this.id == null && uid.id == null ) {<br>result = true;<br>} else if ( this.id == null && uid.id != null ) {<br>result = false;<br>} else if ( this.id != null && uid.id == null ) {<br>result = false;<br>} else {<br>result = Arrays.equals(this.id,uid.id);<br>}<br>}//end if<br>return result;<br>} |
| d. Lazy initialization of "static" fields should be "synchronized" | Dataset 1 | The tool could not identify code patterns. | public static void main(String[] args) {<br>t1 = new Thread(new Runnable() { // issue_line<br>public void run() {<br>increment();<br>}<br>});<br><br>t2 = new Thread(new Runnable() {<br>public void run() { |
| | | | t1.start();<br>// t1 starts after t2. Now, t1's increment might also be called or t2's increment() might also be called. If t2 calls increment(), then the join() method below (you are joining the in the main thread) will be completed and "n" will be printed (if t1 is not getting appropriate time of execution..)<br>increment();<br>}<br>});<br><br>t2.start(); // t2 starts first<br><br>try {<br>t2.join();<br>} catch(InterruptedException e) {<br>e.printStackTrace();<br>}<br><br>System.out.println(n); // increment() might not have been called by t1<br>} |
| e. Magic numbers should not be used | Datasets 1 and 2 | Rules enforcement conflicts with the intended code behaviour. | int[][] one =<br>{ { 1, 2, 3 }, // issue_line<br>{ 4, 5, 6 } }; |
| | | Sufficient readability even without following the rule. | class C285456{<br>public static final String SAMPLE_TEXT = "This is a sample code line 1.\nThis is a sample code line 2.\r\n\nWarm Regards,\r\nSomeUser.";<br><br>public static void main (String[] args) throws java.lang.Exception {<br>String[] lines = SAMPLE_TEXT.split("\\r?\\n"); // catches Windows newlines (\r) as well)<br>for (int i = 0; i < lines.length - 3; i++) { // lines.length - 3 to discard the last 3 lines // issue_line<br>System.out.println(lines[i]);<br>}<br>}<br><br>} |
| f. Member variable visibility should be specified | Dataset 2 | Rules enforcement conflicts with the intended code behaviour. | class constants$4 {<br><br>static final FunctionDescriptor BN_get_rfc3526_prime_6144$FUNC = FunctionDescriptor.of(C_POINTER,<br>C_POINTER<br>);<br>static final MethodHandle BN_get_rfc3526_prime_6144$MH = RuntimeHelper.downcallHandle(<br>openssl_h.LIBRARIES, "BN_get_rfc3526_prime_6144",<br><br>"(Ljdk/incubator/foreign/MemoryAddress;)Ljdk/incubator/foreign/MemoryAddress;",<br>constants$4.BN_get_rfc3526_prime_6144$FUNC, false<br>);<br>static final FunctionDescriptor BN_get_rfc3526_prime_8192$FUNC = FunctionDescriptor.of(C_POINTER,<br>C_POINTER<br>);<br>static final MethodHandle BN_get_rfc3526_prime_8192$MH = RuntimeHelper.downcallHandle(<br>openssl_h.LIBRARIES, "BN_get_rfc3526_prime_8192",<br><br>"(Ljdk/incubator/foreign/MemoryAddress;)Ljdk/incubator/foreign/MemoryAddress;",<br>constants$4.BN_get_rfc3526_prime_8192$FUNC, false<br>);<br>static final FunctionDescriptor ASN1_STRING_length$FUNC = FunctionDescriptor.of(C_INT, // issue_line<br>C_POINTER<br>); |
| g. Source code should be indented consistently | Datasets 1 and 2 | The tool could not identify code patterns. | class RepaintFrameTest extends JFrame {<br><br>private JPanel panel = new JPanel();<br>private JLabel labelOne = new JLabel("label1");<br>private JLabel labelTwo = new JLabel("label2");<br>private JLabel labelThree = new JLabel("label3"); |
| | | | public RepaintFrameTest() {<br>add(panel);<br>}<br><br>public void reapplyLabels() {<br>panel.add(labelOne);<br>panel.add(labelTwo);<br>panel.add(labelThree);<br><br>revalidate();<br>repaint();<br>}<br><br>public static void main(String args[]) {<br>EventQueue.invokeLater(new Runnable() {<br>@Override<br>public void run() {<br>try { // issue_line<br><br>UIManager.setLookAndFeel(UIManager.getSystemLookAndFeelClassName());<br>} catch (ClassNotFoundException \| InstantiationException \| IllegalAccessException \| UnsupportedLookAndFeelException ex) {<br>} |
| h. Track lack of copyright and license headers | Dataset 2 | The tool could not identify code patterns. | **/* // issue_line<br>* Licensed to the Apache Software Foundation (ASF) under one or more<br>* contributor license agreements. See the NOTICE file distributed with<br>* this work for additional information regarding copyright ownership.<br>* The ASF licenses this file to You under the Apache License, Version 2.0<br>* (the "License"); you may not use this file except in compliance with<br>* the License. You may obtain a copy of the License at<br>*<br>* http://www.apache.org/licenses/LICENSE-2.0<br>*<br>* Unless required by applicable law or agreed to in writing, software<br>* distributed under the License is distributed on an "AS IS" BASIS,<br>* WITHOUT WARRANTIES OR CONDITIONS OF ANY KIND, either express or implied.<br>* See the License for the specific language governing permissions and<br>* limitations under the License.<br>*/<br>package jakarta.el;<br><br>import java.lang.reflect.Method;<br>import java.lang.reflect.Modifier; |

## 4.4 RQ4: Are there any gaps in code violation classification between selected SCA tools and established standards?

When annotating warnings generated by SCA tools, we initially investigated whether SCA rules were aligned with the standards. Then, the warnings were labelled according to the rule description if it was entirely aligned with the standards. Otherwise, warnings were labelled based on the defined standards. However, if there were no defined standards, then we labelled warnings only using the rule description. In this process, we identified some gaps on code violation classification between selected SCA tools and established standards are reported in our replication package[28] [17], and summarised in Table 19. Here it is shown that the most frequent gap between defined standards and SCA rule description was the inability to identify any standards for some SCA tool rules. For instance, there were no relevant standards for the SonarQube rule – "Methods should not have too many lines". In fact, such a rule should define/provide a value for the maximum number of lines. However, according to SOLID, a method should only have one responsibility (Single responsibility principle), whereas the Java method should have only one reason to be modified [69]. Even to achieve a single responsibility, the number of lines required in a method body may vary, thus researchers need to examine this issue [70]. Furthermore, we did not identify standards for specific rules, possibly because these standards are still to be established. As a result, we considered this to be a gap between SCA tool developers and standard organisations and these rules were identified as non-standardised rules.

[28] Replication package >> Appendices >> Appendix I_Disagreements on code violation classification between the sele.docx

Another gap is that rules do not conform to the identified standards. For example, according to the SonarQube rule – "An open curly brace should be located at the beginning of a line", the open curly brace should be located at the beginning of the code line in source code. Still, according to the Google Java Coding Style Guide, the open curly brace should be located at the end of the previous code line. Much like the previously mentioned example, two rules for each chosen SCA tool did not align with the current standards.

Table 19 Gaps between SCA rule description and defined standards

| Tool | Gaps between SCA rule description and defined standards | Number of Rules |
|---|---|---|
| PMD | The rule is not adherent to the standard | 2 |
| PMD | The authors could not find any standards for the rule | 41 |
| SonarQube | The rule is not adherent to the standard | 2 |
| SonarQube | The authors could not find any standards for the rule | 67 |

Overall, we believe that the cause of these gaps above lies in the experience and knowledge gap between the Tool Developers and the body responsible for defining coding standards. SCA tools have defined their rules based on the experience of the software engineering community and their preference for coding styles[29]. In contrast, the standards are defined by facts, evidence, and a higher level of experience in software engineering. Consequently, disagreements may arise between the two views. However, academia must undertake research to identify or establish standards for rules that lack defined standards.

# 5 Discussion and Implications

This study focuses on the rules used to detect possible code quality violations by selected SCA tools (PMD and SonarQube), the rules prone to false positives, the reasons for generating false positives, and gaps between SCA tools' rules and established standards. Two source code datasets comprising Stack Overflow and Apache Tomcat code were analysed using PMD and SonarQube to create four violation datasets. Since it is not practical to manually check all the issues reported, four random samples with a 95% confidence level and 5% error rate were prepared for further investigation. We answer our research questions based on the outcomes of this investigation.

**RQ1: What are the rules and standards of SCA tools that generate warnings for specific source code datasets?**

Outcomes show that the majority of PMD rules triggered SCA warnings for the code datasets studied. In comparison, only approximately 50% of SonarQube rules produced SCA warnings for the same datasets. Accordingly, PMD rules were more relevant than SonarQube rules for our four datasets. Moreover, for the rules which generated warnings from both the SCA tools, most of the rules adhere to at least one standard researched in Section 3.4. However, appropriate standards are not seen for many rules. PMD used 12 established standards for their rules, while SonarQube used 11 established standards. Our outcomes here provide a good understanding of the current rules of both PMD and SonarQube and the applicability of established standards for PMD and SonarQube. Although most rules of both SCA tools have defined standards, it is essential to investigate further the SCA rules that lack defined standards.

[29] https://community.sonarsource.com/c/clean-code/43

We compared the outcomes of RQ1 with those of existing studies summarised in Section 3. According to existing studies, Fatima et al. [35], Ramler et al.[36], Ashfaq et al. [37], and Lenarduzzi et al. [39] identified some of the rules available in SCA tools. However, these studies do not identify all the rules available in the SCA tools they examined, and they have not identified any established standards applicable to the rules of the SCA tools. Moreover, Charoenwet et al. [40], Ozturk et al. [41], and Alqaradaghi and Kozsik [42] identified standards for several SCA rules based on CWEs maintained by Mitra and OWASP. However, these studies were also not comprehensive as they studied only a few rules. Furthermore, Bánsághi et al. [43] and Forouzani et al. [44] conducted their study to map rules of SCA tools to the quality characteristics defined in software quality models such as ISO/IEC 9126 and ISO/IEC 25010. However, these efforts were made to categorise the rules in SCA tools, and they do not identify any established standards applicable to the rules of SCA tools. Conversely, our study comprehensively studied all the rules that generated warnings for both the Stack overflow code dataset and the Apache Tomcat code dataset, identifying established standards applicable to these rules. Additionally, we used the established standards to investigate potential gaps between the descriptions of each SCA tool's rules and the standards. The results of this investigation are reported in RQ4.

**Recommendations from RQ1**

The outcome of the RQ1 provides several recommendations for the **Tool Developers**, the **Practitioner Community** and the **Research Community**. **Tool Developers** can follow established standards to define rules. This helps to prevent the Practitioner Community from using non-standardised coding patterns. Moreover, they need to mention the referred standards in the rule description for more transparency of the applicable standards for the SCA tool's rules. To extend the effort taken by the Tool Developers, the **Practitioner Community** also needs to follow established standards when writing source code and fixing SCA warnings. This helps to standardise the coding patterns used in the software. An option to ensure they use only the rules that adhere to established standards is to disable those that do not comply with such standards. Because fixing some warnings (mostly warnings generated by rules that do not follow established standards) might not be necessary and may add more complexity unnecessarily. However, to identify the rules which do not follow establish standards, the **Research Community** should conduct more comprehensive studies on SCA rules and established standards mainly for other widely used SCA tools and rules we haven't considered in this study as we have only considered the rules which generated warnings for our two source code datasets. Furthermore, the Research Community can conduct further studies to find more established standards that we might have missed.

**RQ2: Which SCA rules are prone to false positives?**

Results in the previous section shows that only a few rules produce false positives, and among these, only some account for most false positive warning rates. However, most of the rules that generated false positive warnings have established standards. Moreover, addressing the most frequently generated false positives would decrease the false positive percentage, as those rules contribute 72%, 64%, 83%, and 30% for false positive warning count of our four datasets, respectively.

Zheng et al. [18] used commit histories to label the warning dataset. They classified a warning as a true positive if it was resolved in a subsequent commit and as a false positive if it remained unresolved in a subsequent commit. Wang et al. [24] were also used a similar approach to annotate their warning dataset. Hegedus and Ferenc [10] annotated their warning dataset using "//NOSONAR" comment in the source code. Originally, this comment was used to ignore the warning in SonarQube. Therefore, if the comments were attached to the reported warning, then the warning was labelled as a false positive warning, and warnings were classified as true positives if there were no "//NOSONAR" comments attached to the reported warnings. However, the quality of these automated annotated datasets remains concerning [11] and to annotate warning datasets, these studies did not use any established standards. To address some of these issues, artificially synthesised datasets such as the Juliet vulnerability dataset [12] was created. However, the data points in these datasets exhibit limited diversity. Conversely, our datasets were annotated manually using established standards outlined in the Section 3.4.

**Recommendations from RQ2**

The results generated from RQ2 suggest several recommendations for the **Tool Developers**, the **Practitioner Community** and the **Research Community**. According to the results, **Tool Developers** should fix rules that are frequently prone to false positives for two reasons. The first reason is that identifying false positive warnings wastes developers' time. The second reason is that the Practitioner Community might not identify false positive warnings, and trying to fix these warnings might make the software buggy and lower the quality of the source code. Conversely, the **Practitioner Community** can disable or give more attention to rules that are prone to false positives. This will help the Practitioner Community to use standard coding patterns while preventing them from blindly fixing SCA warnings. To identify the rules that are more prone to false positives, the **Research Community** should study these rules and define standards to fill the gap if there is a lack of standards.

**RQ3: What are the reasons for reported false positives by SCA tools?**

The reasons for SCA tools to generate false positive suggestions are the tools' misidentification of source code patterns, fixing the warnings generated by SCA tools is not required in accordance with the established standards, and the purpose of the source code cannot be fulfilled if the developer follows the warning. Therefore, fixing the rules of SCA tools to mitigate these reasons can reduce the SCA tools from generating false warnings. In contrast, established standards may not be considered newly introduced rules, and with the improvement of those standards, some false warnings may be reclassified as actual warnings.

We conducted a comprehensive literature survey to compare our study with existing studies. According to our findings, there are no related studies that discuss the reasons for generating false positive warnings. To fill this gap, we analysed the false positive warnings according to the established standards to identify the causes behind generating false positive warnings. This streamlines the annotation of warnings in an unlabelled warning dataset. Additionally, it offers insights to SCA Tool Developers, helping them refine the rules that produce a significant amount of false positive warnings, thus lessening the burden on practitioners.

**Recommendations from RQ3**

The results of RQ3 have multiple implications for **Tool Developers**, the **Practitioner Community**, and the **Research Community**. We identified several rules that contribute to the most false positive warnings. By fixing those rules, SCA tools will generate significantly fewer false-positive warnings, encouraging their use. We recommend that **Tool Developers** prioritise fixing those rules that cause false positives generated by SCA tools, by reducing the causes of false positives. Conversely, the **Practitioner Community** can reduce common patterns of causes for generating false positives, if possible, while developing software. Furthermore, to identify more causes of generating false positives, the **Research Community** should study this area and determine how to prevent these causes from both the Tool Developers' and the Practitioner Community's perspectives.

**RQ4: Are there any gaps in code violation classification between selected SCA tools and established standards?**

The most frequent gap between SCA rule description and defined standards was that the authors could not find standards for some rules of SCA tools. There could be two reasons for this. The first reason could be that standards can exist for some of those rules. However, authors might have missed those standards. The second possible reason could be that some rules' standards are yet to be defined, and researchers are still studying those areas. Hence, it would be a valuable research contribution for academia and the software engineering Practitioner Community to identify/define standards for rules that currently lack defined standards. Another gap is that some SCA tools' rules do not adhere to the established standards. In this study, we identified rules which do not follow the established standards. Therefore, the developers of SCA tools or the users of the tools can ignore the rules as they consistently produce false warnings.

To compare the results generated by our study, we have conducted a comprehensive literature study to identify existing studies on gaps between established standards and rule descriptions of SCA tools. However, we were unable to find any related studies. Therefore, the outcome of RQ4 provides a significant contribution to fill this gap.

**Recommendations from RQ4**

The results of RQ4 offer multiple recommendations for **Tool Developers**, the **Practitioner Community**, and the **Research Community**. It is recommended that **Tool Developers** modify SCA tools and their rules to prevent gaps between the description of SCA tools' rules and established code quality standards. In addition, the **Practitioner Community** should also understand the existence of gaps between established standards and SCA tools' rules descriptions and the impossibility of fixing all the warnings generated by SCA tools, as there are false positive warnings. Therefore, the Practitioner Community can disable rules that disagree with established standards or thoroughly analyse warnings generated by these rules. However, these gaps may arise from our failure to find certain standards. Therefore, the **Research Community** should study gaps to find any missing standards. Additionally, they should define standards for the SCA tool's rules that lack established standards.

## 6 Threats to validity

In this section, the potential threats to the validity of this study are discussed [26, 28].

**Internal Validity:** We considered all warnings generated for each code dataset independently, even though they were often generated from the same files or projects. Fixing one issue in one file may automatically fix or introduce other warnings. While this threat is minimal, it was not entirely eliminated, as for most rules, each warning has only a localised impact on the line or block that generated the issue. Some examples of such rules are "CommentRequired" and "MethodArgumentCouldBeFinal" in Dataset 1. As such, the warnings generated by the rule "CommentRequired" do not impact other warnings. To investigate this issue, we examined the top rules that accounted for the most warnings in our dataset. As shown in Appendix C and Appendix D, the most frequently triggered rules, such as "CommentRequired" in PMD, and "Magic numbers should not be used" and "Source code should be indented consistently" in SonarQube, are each tied to a specific, isolated line of code and a single well-defined issue. These rules collectively account for 62% of all PMD warnings and 94% of all SonarQube warnings in our datasets, suggesting that the majority of warnings in our study are indeed independent at the line level. On the other hand, there can be dependency between warnings if SCA tools might identify the same coding pattern as an issue for one rule but not for another. For example, SonarQube rules "An open curly brace should be located at the end of a line" and "An open curly brace should be located at the beginning of a line" contradict each other. Therefore, we used established standards to select a single rule where there were contradictions, preventing such dependence in which the rules generate dependent warnings for the same source code line. In this regard, the likelihood of generating dependent warnings is likely minimal. However, we acknowledge that a small proportion of warnings from more complex rules may be structurally related, and there can be some dependency between different SCA warnings. Therefore, it is worth examining the impact of such rules, and we plan to investigate this aspect in future work, as mentioned in Section 7. Furthermore, we performed several modifications to the Stack Overflow code dataset to make the codebase compilable. By doing so, there is a risk that the code modifications might have contributed to some warnings. However, our goal in this study is to find as many warnings as possible and, if modifying the file increases the number of warnings, that is beneficial to our analysis. Of note is that our focus was aimed at understanding warnings, so even deliberate attempts at generating warnings would have been useful.

Moreover, as explained in the Section 3.4, the standards used and presented in this study were identified. Although most rules have defined standards, a considerable number of rules lack such specifications. In this context, the descriptions of the rules were utilised to identify false positives. Consequently, some standards remain unidentified, and based on these standards, the labels of four datasets may be inaccurate. Nonetheless, we have endeavoured to find all applicable standards to the best of our ability. Warnings were manually assessed according to established standards to label four datasets. Moreover, using the official tool documentation as a guideline to label warnings generated by rules without established standards may introduce a degree of dependency on each tool's internal justification. However, it allowed for consistent interpretation across tools and avoided arbitrary or subjective external judgments for tool-specific rules. Furthermore, this approach encourages readers to identify rules without established standards and to define new standards. Furthermore, despite adherence to the established standards, the authors may inadvertently misclassify the warnings due to misunderstandings. To minimise errors, all authors discussed the standards, and several software tools were developed to assist in the annotation process. Additionally, 10% of Dataset 1 and Dataset 3 were selected randomly, and two authors reviewed these samples together [71-73]. There were three data points out of 115 data points (51 data points in Dataset 1 and 64 data points in Dataset 3) which recorded disagreements at the beginning of the reliability assessment process. The Inter-rater agreement was calculated using Cohen's Kappa between the authors [74], which was found to be 0.6547, indicating a substantial agreement between the authors based on the categories provided by Landis and Koch [75]. However, full agreement on the 115 samples was reached after a round of discussions [74], attesting to the strong reliability of our datasets development [75].

We defined a false positive in this paper as warnings generated by SCA tools that do not align with the defined standards (or the tools' official documentation, if no defined standards exist). Therefore, any warning that deviates from defined standards is excluded and classified as a false positive. However, this approach does not incorporate the preferences of individual software development practitioners or software project teams. For example, there are two specific rules: "An open curly brace should be located at the beginning of a line" and "An open curly brace should be located at the end of a line" in SonarQube which are contradicting, and only the second rule aligns with the established standard. However, there may be a software engineer or team that prefers the first style, where they put curly braces at the start of the code line, and, according to established standards, this is a wrong practice. Therefore, we used multiple standards to ensure consistency. We identified most of the standards from SCA tools' documentation and standards popular in existing studies. Therefore, these standards are popular in the industry and well-defined. Furthermore, we did not find any conflicting standards. To label SCA warnings, we need a well-defined scientific basis, and we identified that following the defined standards for labelling warnings will be appropriate. Therefore, we used the defined standards to label SCA warnings in our analysis to ensure consistency.

**External Validity:** For this study, two source code datasets were analysed using PMD and SonarQube. Only the rules of the SCA tools that generated warnings were examined further. As a result, no additional analysis was conducted on the rules that did not produce warnings. Therefore, the results generated from this study might not be generalizable to false positives generated by other SCA tools. However, most rules generated code quality alarms for the source code, reducing threats. Another threat to the external validity of this study is that only two source code datasets were used, which may not be adequate to generalise the results to other datasets. For instance, the "CommentRequired" PMD rule produced the highest number of warnings for the source code examined. Nevertheless, this rule may not generate the most warnings for another source code dataset. However, as both code datasets exhibited the same pattern, this threat remains minimal. Finally, as explained in Section 4.1 (RQ1), SCA tool's rules change over time. Some new rules may be introduced, while some existing rules may be removed. Consequently, the results of this study may diverge from outcomes for future versions of the SCA tools studied. However, we utilised the latest version of the SCA tool at the time during the study to minimise this threat.

## 7 Summary and Future Work

**Summary**: In this study, we identified several approaches that were taken to create datasets, including real-world program vulnerability datasets and artificially synthesised vulnerability datasets. In addition, other than these two types of vulnerability datasets, manually annotated datasets are also used to train ML models. Moreover, we identified from existing studied that there are no existing studies that comprehensively study SCA tools' rules, source code quality standards applicable for these rules, rules that are prone to false positives, reasons for rules generating false positives and gaps between source code quality standards and descriptions of SCA tools' rules. To fill the gaps identified in existing studies, we created four manually labelled datasets as described in Section 2. Additionally, four datasets were manually annotated using established standards, and any discrepancies between the rules of the SCA tool and the established standards were noted.

According to the outcomes generated for RQ1, SCA tools have defined a large number of rules; however, not all the rules generated warnings for our datasets. PMD generated warnings for 221 rules out of 271, and SonarQube generated warnings for 359 rules out of 632 rules. Additionally, the majority of the rules which generated warnings have established standards, and only a small number (two) of rules are against established standards. Among the rules that generated warnings, we identified 178 with established standards and 41 without standards for PMD. Moreover, in SonarQube, 290 rules had defined standards, and 67 did not, among the rules that generated warnings for two source code datasets. Moreover, among the established standards, the three most used standards are the JDK documentation, the SEI website and the Clean Code book. Tool Developers have classified the rules and presented them on their official websites. We identified that PMD and SonarQube have reported warnings for our datasets covering almost all categories.

From RQ2, we identified the SCA tools' rules that generated false positive warnings, where we observed that PMD mostly generated false positive warnings for the "CommentRequired" rule for both source code datasets, where out of the false positives generated, 72.45% and 64% were generated by this rule in Dataset 1 and Dataset 2, respectively. Moreover, SonarQube generated the majority of false-positive warnings for the "Magic number should not be used" rule across the two datasets, accounting for 83.33% and 30% of all false-positive warnings in Dataset 3 and Dataset 4, respectively. Overall, we identified that only few rules generated false positive warnings.

RQ3 identified the reason for generating false positive warnings, where the majority of false positive warnings may not be necessary based on established standards. Therefore, the Practitioner Community can safely ignore these warnings without compromising the quality of the source code. SonarQube generated false positive warnings mostly for the "Stack overflow" dataset because even without following the warning, the readability of the code was not reduced. Therefore, following the warning is not necessary. Moreover, SonarQube generated false positives for the Apache Tomcat code dataset because SonarQube has not identified code patterns correctly and generated wrong suggestions.

RQ4 identified gaps between warnings of selected SCA tools and established standards. We observed that the most frequent gap was the unavailability of standards for the rules of the SCA tools. Another gap was that the rules do not adhere to established standards.

To improve SCA tools and source code quality, Tool Developers can follow standards to define rules of SCA tools and mention the standards along with the documentation. According to the established standards, rules which generate false positive warnings were identified in this study and Tool Developers can fix these rules that are frequently prone to false positives by reducing the causes of false positives when developing rules for SCA tools. Consequently, after these modifications, gaps in the description of SCA tools' rules and established code quality standards will be settled.

To extend the effort taken by the Tool Developers, the Practitioner Community also needs to follow established standards when writing source code and fixing SCA warnings. They can disable the SCA tool's rules with no established standards and rules that do not comply with such standards. However, even after disabling these rules, there can still be some rules that are prone to false positives, and they may give more attention to these rules. Moreover, source code quality can be improved by disabling rules that disagree with established standards or thoroughly analysing warnings generated by these rules. Likewise, the Practitioner Community can reduce common patterns of causes for generating false positives, if possible, while developing software. Conversely, the Practitioner Community should also understand and be educated the existence of gaps between established standards and SCA tools' rules descriptions, aware that it may be impossible to fix all the warnings generated by SCA tools.

To continue to identify the rules with the above-mentioned issues, the Research Community should conduct more comprehensive studies on SCA rules and established standards, mainly for other widely used SCA tools and rules we haven't considered in this study. In this study, we have only considered the rules which generated warnings for our two source code datasets. Therefore, there are some missing rules that we have not studied and missing rules that may be prone to false positives. To identify these rules, the Research Community should study these rules and define standards to fill the gap if there is a lack of standards. Furthermore, to identify more causes of generating false positives, the Research Community should study this area and identify how to prevent those causes from both the Tool Developers' and the Practitioner Community's perspectives. However, gaps between the description of SCA tools' rules and established code quality standards may arise, given our failure to find certain standards. Therefore, the Research Community should study these gaps to find any missing standards. Additionally, they should define standards for the SCA tool's rules that do not have established standards.

**Future Work**: In this study, we manually labelled only four sample datasets from four larger datasets. The next step is to utilise ML approaches to train the sample datasets and apply the best-performing models to predict the labels for the entire dataset [76]. This will allow us to contribute the four larger datasets to the software engineering community. Our study was conducted for PMD and SonarQube using four sample datasets. However, there are several other SCA tools widely used in the software engineering community. Therefore, it would be beneficial to expand this study to include some of the other frequently used SCA tools. In addition, our study focused only on rules that generated warnings for our two source code datasets. Thus, we plan to conduct a broader study to identify all the rules available in PMD and SonarQube. Expanding this study to include additional rules will enhance our ability to identify more causes of false positives at the individual rule level and discover more existing standards. Moreover, there are constants in the rules. For example, the PMD rule "ExcessiveParameterList" has a constant to define the maximum number of allowed parameters for a Java method. However, established standards do not define this number. Therefore, comprehensively studying the constant values in the rules of SCA tools is a potential significant contribution to the Practitioner Community, Tool Developers, and Research Community. Furthermore, we identified a threat to the validity of the outcomes of this study in Section 6, that there was a possibility of inter-dependencies among multiple SCA warnings within the same class, so that resolving one SCA warning can automatically fix or create other warnings. This issue is to be considered in future work.

# 8 Data Availability

Our replication package is accessible online at https://doi.org/10.5281/zenodo.15770053, thereby facilitating the reproducibility and transparency of this research. Furthermore, the online repository comprises datasets, relevant information on rules and standards, and appendices.

# CRediT authorship contribution statement

**Lakmal Deshapriya**: Conceptualisation, Data curation, Methodology, Writing-Original draft preparation. **Sherlock A. Licorish**: Conceptualisation, Data curation, Supervision, Writing-Reviewing and Editing, Statistics Validation. **Brendon J. Woodford**: Data curation, Supervision, Writing-Reviewing and Editing.

# 9 References

[1] L. Marks, Y. Zou, and A. E. Hassan, "Studying the fix-time for bugs in large open source projects," in Proceedings of the 7th International Conference on Predictive Models in Software Engineering, 2011, pp. 1-8.

[2] S. Meldrum, S. A. Licorish, C. A. Owen, and B. T. R. Savarimuthu, "Understanding stack overflow code quality: A recommendation of caution," *Science of Computer Programming,* vol. 199, pp. 102516, 2020.

[3] V. R. L. de Mendonca, C. L. Rodrigues, F. A. A. de MN Soares, and A. M. R. Vincenzi, "Static analysis techniques and tools: A systematic mapping study," 2013.

[4] F. Cheirdari, and G. Karabatis, "Analyzing False Positive Source Code Vulnerabilities Using Static Analysis Tools," in 2018 IEEE International Conference on Big Data (Big Data), 2018, pp. 4782-4788.

[5] J. Park, I. Lim, and S. Ryu, "Battles with false positives in static analysis of JavaScript web applications in the wild," in 2016 IEEE/ACM 38th International Conference on Software Engineering Companion (ICSE-C), 2016, pp. 61-70.

[6] R. Telang, and S. Wattal, "An empirical analysis of the impact of software vulnerability announcements on firm stock price," *IEEE Transactions on Software engineering,* vol. 33, no. 8, pp. 544-557, 2007.

[7] D. Leffingwell, and D. Widrig, *Managing software requirements: a unified approach*: Addison-Wesley Professional, 2000.

[8] S. H. Kan, *Metrics and models in software quality engineering*: Addison-Wesley Professional, 2003.

[9] A. Tanwar, H. Manikandan, K. Sundaresan, P. Ganesan, S. K. Chandrasekaran, and S. Ravi, "Assessing Validity of Static Analysis Warnings using Ensemble Learning," *arXiv preprint arXiv:2104.11593*, 2021.

[10] P. Hegedus, and R. Ferenc, "Static Code Analysis Alarms Filtering Reloaded: A New Real-World Dataset and its ML-Based Utilization," *IEEE ACCESS,* vol. 10, pp. 55090-55101, 2022.

[11] R. Croft, M. A. Babar, and M. M. Kholoosi, "Data quality for software vulnerability datasets," in 2023 IEEE/ACM 45th International Conference on Software Engineering (ICSE), 2023, pp. 121-133.

[12] T. Boland, and P. E. Black, "Juliet 1. 1 C/C++ and java test suite," *Computer,* vol. 45, no. 10, pp. 88-90, 2012.

[13] Y. Lin, Y. Li, M. Gu, H. Sun, Q. Yue, J. Hu, C. Cao, and Y. Zhang, “Vulnerability dataset construction methods applied to vulnerability detection: A survey,” in 2022 52nd Annual IEEE/IFIP International Conference on Dependable Systems and Networks Workshops (DSN-W), 2022, pp. 141-146.
[14] H. Li, Y. Hao, Y. Zhai, and Z. Qian, “Assisting static analysis with large language models: A chatgpt experiment,” in Proceedings of the 31st ACM Joint European Software Engineering Conference and Symposium on the Foundations of Software Engineering, 2023, pp. 2107-2111.
[15] Y. Tan, and J. Tian, “A Method for Processing Static Analysis Alarms Based on Deep Learning,” *Applied Sciences,* vol. 14, no. 13, pp. 5542, 2024.
[16] D. H. Nguyen, A. Seo, N. P. Nnamdi, and Y. Son, “False Alarm Reduction Method for Weakness Static Analysis Using BERT Model,” *Applied Sciences,* vol. 13, no. 6, pp. 3502, 2023.
[17] L. Deshapriya, S. Licorish, and B. Woodford, "Taking a Closer Look at Warnings Generated by PMD and SonarQube, their Rules and Compliance to Established Coding Standards - Replication Package," Zenodo, 2025.
[18] Y. Zheng, S. Pujar, B. Lewis, L. Buratti, E. Epstein, B. Yang, J. Laredo, A. Morari, and Z. Su, “D2A: A Dataset Built for AI-Based Vulnerability Detection Methods Using Differential Analysis,” in Proceedings - International Conference on Software Engineering, 2021, pp. 111-120.
[19] L. Cabral, B. Miranda, I. Lima, and M. d'Amorim, “RVprio: A tool for prioritizing runtime verification violations,” *Software Testing, Verification and Reliability,* vol. 32, no. 5, pp. e1813, 2022.
[20] T. Aladics, P. Hegedűs, and R. Ferenc, “A Vulnerability Introducing Commit Dataset for Java: An Improved SZZ based Approach,” 2022.
[21] M. Borg, O. Svensson, K. Berg, and D. Hansson, “Szz unleashed: an open implementation of the szz algorithm-featuring example usage in a study of just-in-time bug prediction for the jenkins project,” in Proceedings of the 3rd ACM SIGSOFT International Workshop on Machine Learning Techniques for Software Quality Evaluation, 2019, pp. 7-12.
[22] J. D. Pereira, J. H. Antunes, and M. Vieira, “A Software Vulnerability Dataset of Large Open Source C/C++ Projects,” in Proceedings of IEEE Pacific Rim International Symposium on Dependable Computing, PRDC, 2022, pp. 152-163.
[23] Y. Chen, Z. Ding, L. Alowain, X. Chen, and D. Wagner, “DiverseVul: A New Vulnerable Source Code Dataset for Deep Learning Based Vulnerability Detection,” in ACM International Conference Proceeding Series, 2023, pp. 654-668.
[24] X. Wang, R. Hu, C. Gao, X.-C. Wen, Y. Chen, and Q. Liao, “ReposVul: A Repository-Level High-Quality Vulnerability Dataset,” in Proceedings of the 2024 IEEE/ACM 46th International Conference on Software Engineering: Companion Proceedings, 2024, pp. 472-483.
[25] R. Just, D. Jalali, and M. D. Ernst, “Defects4J: A database of existing faults to enable controlled testing studies for Java programs,” in Proceedings of the 2014 international symposium on software testing and analysis, 2014, pp. 437-440.
[26] Y. Yang, M. Wen, X. Gao, Y. Zhang, and H. Sun, “Reducing false positives of static bug detectors through code representation learning,” in 2024 IEEE International Conference on Software Analysis, Evolution and Reengineering (SANER), 2024, pp. 681-692.
[27] K. Liu, D. Kim, T. F. Bissyandé, S. Yoo, and Y. Le Traon, “Mining Fix Patterns for FindBugs Violations,” *IEEE TRANSACTIONS ON SOFTWARE ENGINEERING,* vol. 47, no. 1, pp. 165-188, JAN 1, 2021.

[28] S. Yerramreddy, A. Mordahl, U. Koc, S. Y. Wei, J. S. Foster, M. Carpuat, and A. A. Porter, "An empirical assessment of machine learning approaches for triaging reports of static analysis tools," *EMPIRICAL SOFTWARE ENGINEERING,* vol. 28, no. 2, MAR, 2023.
[29] A. Nagaraj, B. Sinha, M. Sood, Y. Mathur, S. Gupta, and D. Sitaram, "Learning Algorithms in Static Analysis of Web Applications," *arXiv preprint arXiv:2210.07465*, 2022.
[30] A. Kharkar, R. Z. Moghaddam, M. Jin, X. Liu, X. Shi, C. Clement, and N. Sundaresan, "Learning to Reduce False Positives in Analytic Bug Detectors," in Proceedings - International Conference on Software Engineering, 2022, pp. 1307-1316.
[31] T. T. Vu, and H. D. Vo, "Using Multiple Code Representations to Prioritize Static Analysis Warnings," in Proceedings - International Conference on Knowledge and Systems Engineering, KSE, 2022.
[32] K.-T. Ngo, D.-T. Do, T.-T. Nguyen, and H. D. Vo, "Ranking warnings of static analysis tools using representation learning," in 2021 28th Asia-Pacific Software Engineering Conference (APSEC), 2021, pp. 327-337.
[33] U. Koc, S. Wei, J. S. Foster, M. Carpuat, and A. A. Porter, "An empirical assessment of machine learning approaches for triaging reports of a Java static analysis tool," in Proceedings - 2019 IEEE 12th International Conference on Software Testing, Verification and Validation, ICST 2019, 2019, pp. 288-299.
[34] S. Lee, S. Hong, J. Yi, T. Kim, C. J. Kim, S. Yoo, and Ieee, "Classifying False Positive Static Checker Alarms In Continuous Integration Using Convolutional Neural Networks," in 2019 IEEE 12TH CONFERENCE ON SOFTWARE TESTING, VALIDATION AND VERIFICATION (ICST 2019), 2019, pp. 391-401.
[35] A. Fatima, S. Bibi, and R. Hanif, "Comparative study on static code analysis tools for c/c++," in 2018 15th International Bhurban Conference on Applied Sciences and Technology (IBCAST), 2018, pp. 465-469.
[36] R. Ramler, M. Moser, and J. Pichler, "Automated static analysis of unit test code," in 2016 IEEE 23rd International Conference on Software Analysis, Evolution, and Reengineering (SANER), 2016, pp. 25-28.
[37] Q. Ashfaq, R. Khan, and S. Farooq, "A comparative analysis of static code analysis tools that check java code adherence to java coding standards," in 2019 2nd international conference on communication, computing and digital systems (C-CODE), 2019, pp. 98-103.
[38] J. Novak, and A. Krajnc, "Taxonomy of static code analysis tools," in The 33rd international convention MIPRO, 2010, pp. 418-422.
[39] V. Lenarduzzi, F. Pecorelli, N. Saarimaki, S. Lujan, and F. Palomba, "A critical comparison on six static analysis tools: Detection, agreement, and precision," *Journal of Systems and Software,* vol. 198, pp. 111575, 2023.
[40] W. Charoenwet, P. Thongtanunam, V.-T. Pham, and C. Treude, "An empirical study of static analysis tools for secure code review," in Proceedings of the 33rd ACM SIGSOFT International Symposium on Software Testing and Analysis, 2024, pp. 691-703.
[41] O. S. Ozturk, E. Ekmekcioglu, O. Cetin, B. Arief, J. Hernandez-Castro, and Acm, "New Tricks to Old Codes: Can AI Chatbots Replace Static Code Analysis Tools?," in PROCEEDINGS OF THE 2023 EUROPEAN INTERDISCIPLINARY CYBERSECURITY CONFERENCE, EICC 2023, 2023, pp. 13-18.
[42] M. Alqaradaghi, and T. Kozsik, "Comprehensive Evaluation of Static Analysis Tools for Their Performance in Finding Vulnerabilities in Java Code," *IEEE Access*, 2024.

[43] A. Bánsághi, B. G. Ézsiás, A. Kovács, and A. Tátrai, “Source code scanners in software quality management and connections to international standards,” in Annales Univ. Sci. Budapest. Sect. Comput, 2012, pp. 81-92.
[44] S. Forouzani, Y. K. Chiam, and S. Forouzani, “Method for assessing software quality using source code analysis,” in Proceedings of the Fifth International Conference on Network, Communication and Computing, 2016, pp. 166-170.
[45] Y. Wang, B. Zheng, and H. Huang, “Complying with Coding Standards or Retaining Programming Style: A Quality Outlook at Source Code Level,” *J. Softw. Eng. Appl.,* vol. 1, no. 1, pp. 88-91, 2008.
[46] C. Boogerd, and L. Moonen, “Assessing the value of coding standards: An empirical study,” in 2008 IEEE International conference on software maintenance, 2008, pp. 277-286.
[47] S. Tandon, V. Kumar, and V. Singh, “Empirical evaluation of code smells in open-source software (OSS) using Best Worst Method (BWM) and TOPSIS approach,” *International Journal of Quality & Reliability Management,* vol. 39, no. 3, pp. 815-835, 2022.
[48] V. Piantadosi, S. Scalabrino, and R. Oliveto, “Fixing of security vulnerabilities in open source projects: A case study of apache http server and apache tomcat,” in 2019 12th IEEE Conference on software testing, validation and verification (ICST), 2019, pp. 68-78.
[49] S. Kaur, and S. Singh, “Prediction Model to Investigate Influence of Code Smells on Metrics in Apache Tomcat,” *International Journal of Advanced Research in Computer Science,* vol. 8, no. 5, 2017.
[50] S. Ganesh, F. Palma, and T. Olsson, “Are source code metrics “Good Enough” in predicting security vulnerabilities?,” *Data,* vol. 7, no. 9, pp. 127, 2022.
[51] S. Ganesh, "Predicting security vulnerabilities using software code metrics," 2021.
[52] T.-Y. Chong, V. Anu, and K. Z. Sultana, “Using software metrics for predicting vulnerable code-components: A study on java and python open source projects,” in 2019 ieee international conference on computational science and engineering (cse) and ieee international conference on embedded and ubiquitous computing (euc), 2019, pp. 98-103.
[53] S. A. Licorish, and M. Wagner, “Combining gin and pmd for code improvements,” in Proceedings of the Genetic and Evolutionary Computation Conference Companion, 2022, pp. 790-793.
[54] S. A. Licorish, and T. Nishatharan, “Contextual profiling of stack overflow java code security vulnerabilities initial insights from a pilot study,” in 2021 IEEE 21st International Conference on Software Quality, Reliability and Security Companion (QRS-C), 2021, pp. 1060-1068.
[55] S. A. Licorish, and M. Wagner, “Dissecting copy/delete/replace/swap mutations: Insights from a gin case study,” in Proceedings of the Genetic and Evolutionary Computation Conference Companion, 2022, pp. 1940-1945.
[56] M. Idris, I. Syarif, and I. Winarno, “Development of vulnerable web application based on OWASP API security risks,” in 2021 International Electronics Symposium (IES), 2021, pp. 190-194.
[57] K. A. Sedek, N. Osman, M. N. Osman, and J. H. Kamaruzaman, “Developing a Secure Web Application Using OWASP Guidelines,” *Comput. Inf. Sci.,* vol. 2, no. 4, pp. 137-143, 2009.
[58] K. Ljung, and J. Gonzalez-Huerta, ““to clean code or not to clean code” a survey among practitioners,” in International Conference on Product-Focused Software Process Improvement, 2022, pp. 298-315.

[59] P. Becker, L. A. Olsina Santos, and M. F. Papa, "Strategy for improving source code compliance to a style guide," in XXVIII Congreso Argentino de Ciencias de la Computación (CACIC)(La Rioja, 3 al 6 de octubre de 2022), 2023.

[60] P. Becker, L. Olsina, and M. F. Papa, "Approach to Improving Java Source Code Considering Non-compliance with a Java Style Guide," in Argentine Congress of Computer Science, 2022, pp. 123-139.

[61] A. Honkaranta, T. Leppänen, and A. Costin, "Towards practical cybersecurity mapping of stride and cwe—a multi-perspective approach," in 2021 29th Conference of Open Innovations Association (FRUCT), 2021, pp. 150-159.

[62] S. Simonetto, and P. Bosch, "Comprehensive threat analysis and systematic mapping of CVEs to MITRE framework," in 1st International Conference on Natural Language Processing and Artificial Intelligence for Cyber Security, NLPAICS 2024, 2024.

[63] V. Arnaoudova, M. Di Penta, and G. Antoniol, "Linguistic antipatterns: What they are and how developers perceive them," *Empirical Software Engineering,* vol. 21, pp. 104-158, 2016.

[64] L. Murphy, T. Alliyu, A. Macvean, M. B. Kery, and B. A. Myers, "Preliminary analysis of REST API style guidelines," *Ann Arbor,* vol. 1001, no. 48109, pp. 4, 2017.

[65] J. Balayla, "Gaussian Distribution, Confidence Intervals, Binomial (Wald) Proportion, Wilson Score Interval and Delta Theorem," *Theorems on the Prevalence Threshold and the Geometry of Screening Curves: A Bayesian Approach to Clinical Decision-Making*, pp. 173-185: Springer, 2024.

[66] E. Zolduoarrati, S. A. Licorish, and N. Stanger, "Impact of individualism and collectivism cultural profiles on the behaviour of software developers: A study of stack overflow," *Journal of Systems and Software,* vol. 192, pp. 111427, 2022.

[67] F. Faul, E. Erdfelder, A.-G. Lang, and A. Buchner, "G* Power 3: A flexible statistical power analysis program for the social, behavioral, and biomedical sciences," *Behavior research methods,* vol. 39, no. 2, pp. 175-191, 2007.

[68] W. G. Cochran, *Sampling techniques*: john wiley & sons, 1977.

[69] R. C. Martin, *Clean Code: A Handbook of Agile Software Craftsmanship*: Prentice Hall PTR, 2008.

[70] E. Tempero, and P. Ralph, "A framework for defining coupling metrics," *Science of Computer Programming,* vol. 166, pp. 214-230, 2018.

[71] K. A. Neuendorf, *The content analysis guidebook*: sage, 2017.

[72] A. Jordan, D. Kunkel, J. Manganello, and M. Fishbein, *Media messages and public health*: New York: Routledge, 2009.

[73] E. Zolduoarrati, S. A. Licorish, and N. Stanger, "Harmonising contributions: exploring diversity in software engineering through CQA mining on stack overflow," *ACM transactions on software engineering and methodology,* vol. 33, no. 7, pp. 1-54, 2024.

[74] J. Cohen, "A coefficient of agreement for nominal scales," *Educational and psychological measurement,* vol. 20, no. 1, pp. 37-46, 1960.

[75] J. R. Landis, and G. G. Koch, "The measurement of observer agreement for categorical data," *biometrics*, pp. 159-174, 1977.

[76] S. R. Addula, and G. S. Sajja, "Automated machine learning to streamline data-driven industrial application development." pp. 1-4.

# Appendices

## Appendix A Number of rules for standards in PMD

| Standards | Number of Rules |
|---|---|
| JDK Doc | 87 |
| https://wiki.sei.cmu.edu | 25 |
| Book - Clean code | 24 |
| Book - Effective Java by Joshua Bloch | 21 |
| https://cwe.mitre.org | 13 |
| Book - Refactoring: Improving the Design of Existing Code | 13 |
| Google Java Coding Style Guide | 11 |
| SOLID principles | 2 |
| https://owasp.org | 1 |
| Michele Lanza and Radu Marinescu. Object-Oriented Metrics in Practice: Using Software Metrics to Characterize, Evaluate, and Improve the Design of Object-Oriented Systems. Springer, Berlin, 1 edition, October 2006. Page 80 | 1 |
| Paper - Linguistic Antipatterns - What They Are and How Developers Perceive Them | 1 |
| Principle of Least Astonishment | 1 |

## Appendix B Number of rules for standards in SonarQube

| Standards | Number of Rules |
|---|---|
| JDK Doc | 120 |
| https://wiki.sei.cmu.edu | 73 |
| https://cwe.mitre.org | 32 |
| Google Java Coding Style Guide | 32 |
| Book - Effective Java by Joshua Bloch | 22 |
| Book - Clean Code | 15 |
| Book - Mastering Regular Expressions by Jeffrey Friedl | 11 |
| Book - Refactoring: Improving the Design of Existing Code | 8 |
| SOLID principles | 5 |
| https://owasp.org | 2 |
| transitive property of equality | 1 |

## Appendix C PMD rules prone to false positives

| Rule | FP Count | | Sample Count | | FP Percentage | |
|---|---|---|---|---|---|---|
| | dataset1 | dataset2 | dataset1 | dataset2 | dataset1 | dataset2 |
| CommentRequired | 71 | 46 | 105 | 52 | 67.62% | 88.46% |
| LawOfDemeter | 9 | 7 | 10 | 45 | 90.00% | 15.56% |
| UseUtilityClass | 8 | 0 | 12 | 0 | 66.67% | 0.00% |
| SingletonClassReturningNewInstance | 1 | 1 | 1 | 1 | 100.00% | 100.00% |
| UnusedAssignment | 1 | 0 | 2 | 0 | 50.00% | 0.00% |
| AvoidArrayLoops | 1 | 0 | 1 | 0 | 100.00% | 0.00% |
| CloneMethodMustBePublic | 1 | 0 | 1 | 0 | 100.00% | 0.00% |
| DoNotUseThreads | 1 | 0 | 1 | 0 | 100.00% | 0.00% |
| DontUseFloatTypeForLoopIndices | 1 | 0 | 1 | 0 | 100.00% | 0.00% |
| GuardLogStatement | 1 | 0 | 1 | 0 | 100.00% | 0.00% |
| JUnit4TestShouldUseAfterAnnotation | 1 | 0 | 1 | 0 | 100.00% | 0.00% |
| JUnit4TestShouldUseTestAnnotation | 1 | 0 | 1 | 0 | 100.00% | 0.00% |
| SingularField | 1 | 0 | 1 | 0 | 100.00% | 0.00% |
| UselessParentheses | 0 | 2 | 0 | 3 | 0.00% | 66.67% |
| AvoidDeeplyNestedIfStmts | 0 | 1 | 0 | 1 | 0.00% | 100.00% |
| AvoidThrowingNewInstanceOfSameException | 0 | 1 | 0 | 1 | 0.00% | 100.00% |
| CloneMethodMustImplementCloneable | 0 | 1 | 0 | 1 | 0.00% | 100.00% |
| DataClass | 0 | 1 | 0 | 1 | 0.00% | 100.00% |
| DetachedTestCase | 0 | 1 | 0 | 1 | 0.00% | 100.00% |
| ImplicitSwitchFallThrough | 0 | 1 | 0 | 1 | 0.00% | 100.00% |
| NonThreadSafeSingleton | 0 | 1 | 0 | 1 | 0.00% | 100.00% |
| ProperLogger | 0 | 1 | 0 | 1 | 0.00% | 100.00% |
| TestClassWithoutTestCases | 0 | 1 | 0 | 1 | 0.00% | 100.00% |
| UnnecessaryLocalBeforeReturn | 0 | 1 | 0 | 1 | 0.00% | 100.00% |
| CloseResource | 0 | 2 | 0 | 2 | 0.00% | 100.00% |
| JUnitTestsShouldIncludeAssert | 0 | 4 | 0 | 4 | 0.00% | 100.00% |
| Total FP | 98 | 72 | | | | |

# Appendix D SonarQube rules prone to false positives

| Rule code | Rule | FP Count | | Sample Count | | FP Percentage | |
|---|---|---|---|---|---|---|---|
| | | **dataset3** | **dataset4** | **dataset3** | **dataset4** | **dataset3** | **dataset4** |
| java:S109 | Magic numbers should not be used | 25 | 13 | 44 | 21 | 56.82% | 61.90% |
| java:S1451 | Track lack of copyright and license headers | 0 | 9 | 0 | 9 | 0.00% | 100.00% |
| java:S1120 | Source code should be indented consistently | 2 | 6 | 105 | 252 | 1.90% | 2.38% |
| java:S2039 | Member variable visibility should be specified | 0 | 6 | 0 | 6 | 0.00% | 100.00% |
| java:S2097 | equals(Object obj) should test the argument's type | 1 | 1 | 1 | 1 | 100.00% | 100.00% |
| java:S5738 | @Deprecated code marked for removal should never be used | 0 | 1 | 0 | 1 | 0.00% | 100.00% |
| java:S3012 | Arrays and lists should not be copied using loops | 0 | 1 | 0 | 1 | 0.00% | 100.00% |
| java:S1258 | Classes and enums with private members should have a constructor | 0 | 1 | 0 | 1 | 0.00% | 100.00% |
| java:S1182 | Classes that override "clone" should be "Cloneable" and call "super.clone()" | 0 | 1 | 0 | 1 | 0.00% | 100.00% |
| java:S2440 | Classes with only "static" methods should not be instantiated | 0 | 1 | 0 | 1 | 0.00% | 100.00% |
| java:S1107 | Close curly brace and the next "else", "catch" and "finally" keywords should be located on the same line | 0 | 1 | 0 | 1 | 0.00% | 100.00% |
| java:S1201 | equals method overrides should accept "Object" parameters | 0 | 1 | 0 | 1 | 0.00% | 100.00% |
| java:S1994 | for loop increment clauses should modify the loops' counters | 0 | 1 | 0 | 1 | 0.00% | 100.00% |
| java:S1696 | NullPointerException should not be caught | 0 | 1 | 0 | 1 | 0.00% | 100.00% |
| java:S1820 | Classes should not have too many fields | 1 | 0 | 1 | 0 | 100.00% | 0.00% |
| java:S2444 | Lazy initialization of "static" fields should be "synchronized" | 1 | 0 | 1 | 0 | 100.00% | 0.00% |
| Total FP | | 30 | 44 | | | | |